\documentclass[%
reprint,
superscriptaddress,
amsmath,amssymb,
aps,prl
]{revtex4-2}

\usepackage{graphicx}
\usepackage{dcolumn}
\usepackage{bm}
\usepackage{hyperref}
\usepackage{float}
\usepackage{cases}
\usepackage{cleveref}
\usepackage{wasysym}
\usepackage[makeroom]{cancel}
\usepackage{scalerel}
\usepackage{lipsum}

\renewcommand{\bm}[1]{\boldsymbol{\mathbf{#1}}}

\providecommand*{\bm}{\mathbf}

\providecommand*{\bcdot}{\bm{\cdot}}

\providecommand*{\bnabla}{\bm{\nabla}}

\renewcommand{\O}{\mathcal{O}}
\hypersetup{
	colorlinks,
	linkcolor={red},
	citecolor={blue},
	urlcolor={black}
}

\begin{document}
	
	\preprint{APS/123-QED}
	\sloppy
	\allowdisplaybreaks
	\title{An unrecognized force in inertial microfluidics}
	
	\author{Siddhansh Agarwal}
	\thanks{S.A. and F.K.C. contributed equally to this work.}
	\author{Fan Kiat Chan}
	\thanks{S.A. and F.K.C. contributed equally to this work.}
	\affiliation{Mechanical Science and Engineering, University of Illinois, Urbana-Champaign, Illinois 61801, USA}
	
	\author{Bhargav Rallabandi}
	\affiliation{Mechanical Engineering, University of California, Riverside, USA}
	
	\author{Mattia Gazzola}
	\affiliation{Mechanical Science and Engineering, University of Illinois, Urbana-Champaign, Illinois 61801, USA}
	\affiliation{National Center for Supercomputing Applications, University of Illinois, Urbana-Champaign, Illinois 61801, USA}
	\affiliation{Carl R.\ Woese Institute for Genomic Biology, University of Illinois, Urbana-Champaign, Illinois 61801, USA}
	
	\author{Sascha Hilgenfeldt}%
	\email{sascha@illinois.edu}
	\affiliation{Mechanical Science and Engineering, University of Illinois, Urbana-Champaign, Illinois 61801, USA}%
	
	
	
	
	\date{\today}
	
	\begin{abstract}
	\end{abstract}
	
	\maketitle
	
	
	{\bf Describing effects of small but finite inertia on suspended particles is a fundamental fluid dynamical problem that has never been solved in full generality \cite{ho1974inertial,lovalenti1993hydrodynamic,schonberg1989inertial,di2009inertial,hood2015inertial}. Modern microfluidics has turned this academic problem into a practical challenge through the use of high-frequency ($\omega\!\!\!\sim$\,kHz--MHz) oscillatory flows, perhaps the most efficient way to take advantage of inertial effects at low Reynolds numbers, to precisely manipulate particles, cells and vesicles without the need for charges or chemistry \cite{lutz2006hydrodynamic,rogers2011selective,thameem2017fast}. The theoretical understanding of flow forces on particles has so far hinged on the pioneering work of Maxey and Riley  (MR in the following) \cite{maxey1983equation}, almost 40 years ago. We demonstrate here theoretically and computationally that oscillatory flows exert previously unexplained, significant and persistent forces, that these emerge from a combination of particle inertia and spatial flow variation, and that they can be quantitatively predicted through a generalization of MR.}

	Oscillatory microfluidics is usually set up by or past a localized object (e.g.\ a microbubble or a no-slip solid \cite{rogers2011selective,lutz2005microscopic}), resulting in spatially non-uniform flows characterized by strong variations on gradient $L_\Gamma$ and curvature $L_\kappa$ length scales. Such flows exert remarkably consistent and controllable forces on particles, and have been employed with great success for guidance, separation, aggregation, and sorting \cite{wang2011size,schmid2014sorting,chen2014manipulation,park2016chip,thameem2016particle,thameem2017fast,volk2020size}. Nonetheless, it is precisely this use of localized oscillations in modern microfluidics that is now pushing the envelope of the MR equation, exposing its limits through the observation of unexplained, significant and persistent forces.  Here we provide a thorough revision of its theoretical foundations, but first, in light of the importance of this work for applications,
	we state a major practical outcome: in any oscillatory background flow field $\bar{\bm{U}}$ associated with a localized object, density-matched ($\rho$) spherical particles (radius $a_p$) experience an attractive force towards the object. The component of this force along the object-to-particle connector $\bm{e}$ takes the explicit form
	\begin{align}
		F_{\Gamma\kappa} = m_f \left\langle a_p^2\nabla \bar{\bm{U}}:\nabla \nabla \bar{\bm{U}}\right\rangle \mathcal{F}(\lambda) \cdot \bm{e}\,,
		\label{dimforce}
	\end{align}
	where $m_f$ is the displaced fluid mass ($=4\pi \rho a_p^3/3$) and the inner product represents the interaction of flow gradients and curvatures. Force (\ref{dimforce}) is steady,  resulting from a time average $\langle\cdot\rangle$. The effect of oscillation frequency is quantified by the universal, analytically derived function $\mathcal{F}$ of the Stokes number $\lambda$. For harmonic oscillatory flows, $\lambda\equiv a_p^2\omega/(3\nu)$ and to excellent approximation $\mathcal{F}(\lambda)$ reads
	\begin{equation}
		\mathcal{F}(\lambda) = \frac{1}{3}+\frac{9}{16}\sqrt{\frac{3}{2\lambda}},
		\label{twotermforce}
	\end{equation}
	valid over {\em the entire range} from the viscous $\lambda\ll 1$ to the inviscid $\lambda\gg 1$ limits. In practice, (\ref{dimforce}) moves a particle against its Stokes mobility along a radial coordinate measuring distance $r_p$ from the localized object, so that the steady equation of motion becomes simply
	\begin{align}
		\frac{d r_{p}}{dt} = \frac{F_{\Gamma\kappa}}{6\pi a_p \nu \rho} \,,
		\label{particleeom}
	\end{align}
	with $\nu$ the kinematic viscosity of the fluid. Generally, $F_{\Gamma\kappa}<0$,  since the amplitude of $\bar{\bm U}$ decays with distance from the oscillating object, indicating attraction. If an additional steady flow component is present, \eqref{particleeom} quantifies the deviation between particle and fluid motion.
	
	\begin{figure*}
		\centering
		\includegraphics[width=\textwidth]{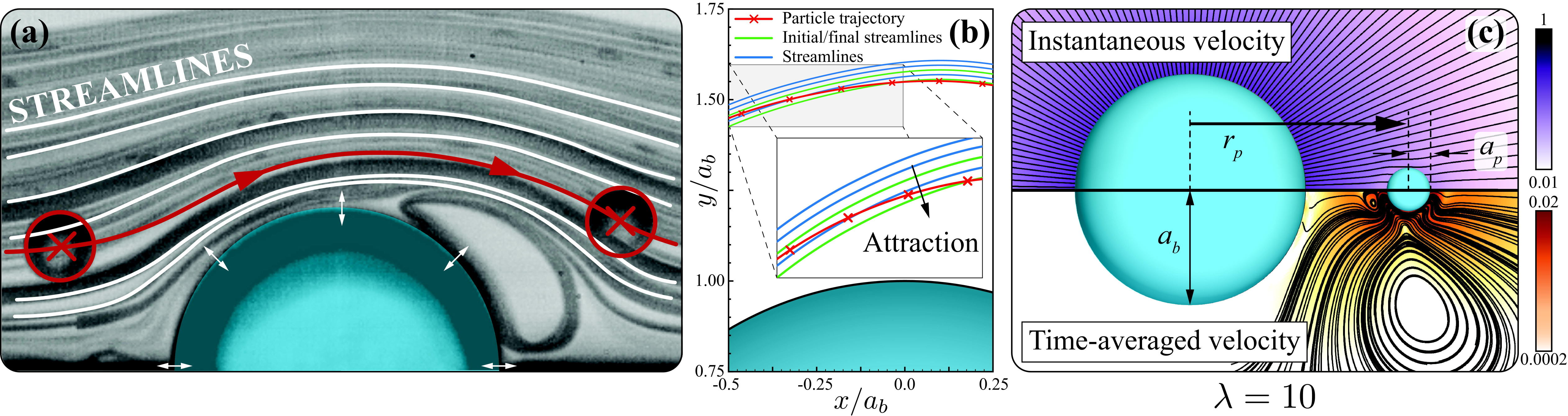}
		\caption{Particle attraction to oscillating bubbles. (a) A neutrally buoyant particle ($a_p=10\mu m$, $\lambda\approx 4$) is transported past an oscillating microbubble ($a_b=40\mu$m, $\omega/(2\pi)=20$kHz). (b) Close-up shows the particle trajectory (red) intersecting streamlines (blue), indicating a net attraction towards the bubble over fast time scales of a few ms, unexplained by existing theories: Inertial particle migration due to shear gradients \cite{di2007continuous,di2009inertial,warkiani2015malaria} is far slower; the secondary radiation force of acoustofluidics \cite{bjerknes1906fields,hay2009model,coakley1978cavitation,bruus2012acoustofluidics,schmid2014sorting} is proportional to the particle-fluid density contrast and thus vanishes here; an ad hoc theory for nearly inviscid flows ($\lambda\gg1$) from  \cite{agarwal2018inertial}  predicts an attraction much too weak to explain observations.
			(c) Simulation of the prototypical problem: a particle exposed to the flow of a bubble oscillating in volume mode at relative amplitude $\epsilon$.
			Top figure: instantaneous streamlines (color bar is flow speed in units of $U^*$); bottom figure: time-averaged streamlines (color bar is steady flow speed in units of $\epsilon U^*$).}
		\label{figure1}
	\end{figure*}
	
	The above equations completely describe the particle dynamics and stem from a rigorous, general formalism developed here to respond to discrepancies observed experimentally. As illustrated in Fig.~\ref{figure1}ab, when neutrally buoyant particles of moderate $\lambda$ approach the surface of oscillating bubbles (cf.\ \cite{wang2012efficient,wang2013frequency,thameem2016particle,thameem2017fast}), we find evidence of significant radial attractive forces, even at a considerable distance from the bubble. This observation is not explained by existing theories  \cite{di2007continuous, warkiani2015malaria,rogers2011selective,hashmi2012oscillating,chen2014manipulation,chen2016onset,park2016chip,bjerknes1906fields,hay2009model,coakley1978cavitation,bruus2012acoustofluidics,schmid2014sorting,agarwal2018inertial} that either predict no attraction at all or a much too weak effect (see caption of Fig.~\ref{figure1}).
	
	Our goal here is to develop a unifying theory that explains observations, accounts for particle inertia, and seamlessly spans the full viscous-to-inviscid operational flow spectrum.
	Accordingly, we revisit MR \cite{maxey1983equation} and systematically account for all leading-order terms in particle Reynolds number $\operatorname{Re}_{p}=a_p U^*/\nu$, with $U^*$ the velocity scale of the background flow. We then reveal their effect through a specially constructed case: a bubble of radius $a_b$ oscillating in pure volume (breathing) mode, with a spherical, neutrally buoyant particle placed at an initial center-to-center distance $r_{p}(0)$. This scenario induces no rectified (streaming) flow in the absence of the particle \cite{lon98}, and therefore allows for the precise evaluation of the newly considered disturbance flow effects introduced by the particle itself. The analysis is complemented by
	direct numerical simulations (DNS) that provide first-principle solutions of flow field and particle displacement.  Figure~\ref{figure1}c (upper half) shows that the computed oscillatory flow component closely resembles the background flow even in the presence of the particle, while time-averaging over an oscillation cycle (bottom half) reveals the much richer secondary steady disturbance flow induced by the particle.
	
	
	Like MR, we wish to describe the hydrodynamic forces on a particle centered at $\bm{r}_p$ using only information from the given undisturbed background flow $\bm{\bar{U}}$.
	We fix
	a (moving) coordinate system at $\bm{r}_p$ and non-dimensionalize lengths by $a_p$, times by $\omega^{-1}$, and velocities by $U^*$ (using lowercase letters for non-dimensional velocities).  A spherical particle exposed to a known (lab-frame) background flow $\bar{\bm{u}}$ and moving with velocity $\bm{u}_p$ (neglecting effects of rotation) then experiences the effects of the undisturbed flow $\boldsymbol{w}^{(0)}=\bm{\bar{u}}-\bm{u}_p$ and a disturbance flow $\boldsymbol{w}^{(1)}$. Following \cite{maxey1983equation}, the latter obeys
	\begin{align}
		&\nabla^{2} \boldsymbol{w}^{(1)}-\nabla p^{(1)} =3\lambda \frac{\partial \boldsymbol{w}^{(1)}}{\partial t}+\operatorname{Re}_{p}\bm{f},\quad {\rm where} \label{distflow}\\
		&\bm{f}=\boldsymbol{w}^{(0)} \cdot \nabla \boldsymbol{w}^{(1)}+\boldsymbol{w}^{(1)}\cdot \nabla \boldsymbol{w}^{(0)} +\boldsymbol{w}^{(1)}\cdot \nabla\boldsymbol{w}^{(1)}\nonumber
	\end{align}
	with boundary conditions $\boldsymbol{w}^{(1)} =\bm{u}_p-\bm{\bar{u}} \, \text { on } r=1$, and $\boldsymbol{w}^{(1)} =0 \, \text { as } r \rightarrow \infty$.
	This equation is exact and does not rely on small $\operatorname{Re}_p$. To obtain explicit results, we use two expansions: one, like MR, expands the background flow around the particle position into spatial moments of alternating symmetry:
	\begin{align}
		\bm{\bar{u}}=\bm{\bar{u}}|_{\bm{r}_{p}} + \bm{r}\cdot \bm{E} + \bm{r}\bm{r}:\bm{G}+\dots, \label{ubarexp}
	\end{align}
	where $\bm{E}=(a_p/L_\Gamma)\nabla \bm{\bar{u}}|_{\bm{r}_{p}}$ and $\bm{G}=\frac{1}{2}(a_p^2/L_\kappa^2)\nabla\nabla \bm{\bar{u}}|_{\bm{r}_{p}}$ capture the background flow shear gradients and curvatures, whose scales are, in practice, much larger than $a_p$, justifying (\ref{ubarexp}).

	\begin{figure*}
		\centering
		\includegraphics[width=\textwidth]{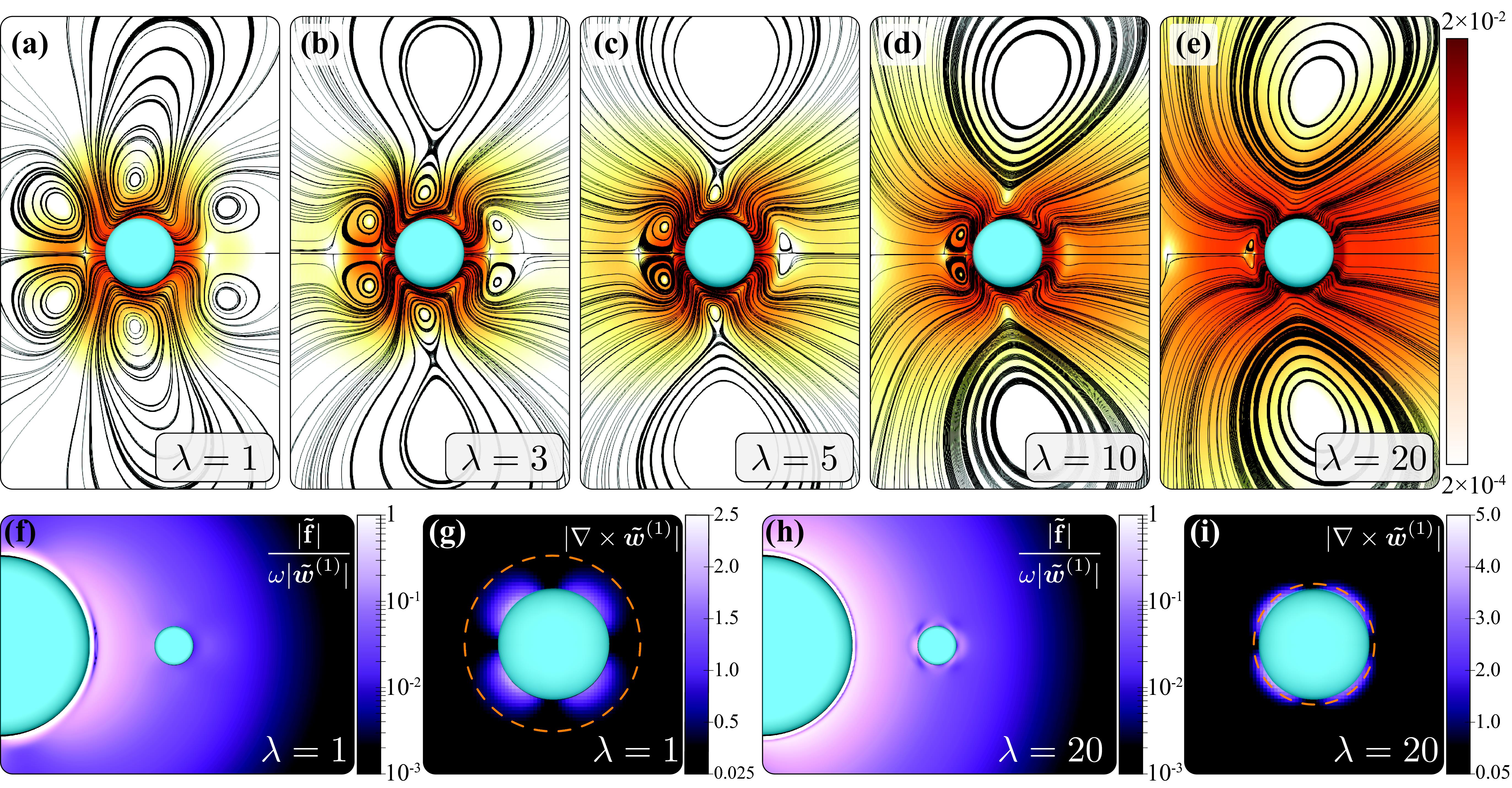}
		\caption{Flow field simulation results. (a-e) Streamlines of the steady flow $\langle \boldsymbol{w}\rangle =  \langle \boldsymbol{w}^{(1)}\rangle$ (Stokes streamfunction isolines) for different $\lambda$; color bar is velocity magnitude in units of $\epsilon U^*$; (f,h) The magnitude of Fourier-transformed quantities (indicated by tildes) evaluated at the driving frequency $\omega$ demonstrates that the flow field has no outer, inertia-dominated region. The ratio between oscillatory disturbance flow advective force ${\bf \tilde{f}(\omega)}$ and the Fourier component of the unsteady inertia $\partial\boldsymbol{{w}}^{(1)}\!\!/\partial t$  remains small away from the bubble. (g,i) The Fourier component of vorticity at $\omega$ is confined to the oscillatory Stokes layer thickness $\delta_S=\sqrt{2\nu/\omega}$ (orange-dashed circle) around the particle.} \label{figure2}
	\end{figure*}
	
	
	The other cornerstone of our theory is
	a regular perturbation expansion of all variables in (\ref{distflow}), using subscripts for orders of $\operatorname{Re}_p$ , e.g., $\boldsymbol{w}^{(1)}=\boldsymbol{w}^{(1)}_0 + \operatorname{Re}_p \boldsymbol{w}^{(1)}_1 + {\cal O}(\operatorname{Re}_p^2)$. In contrast to MR, this retains a term $\operatorname{Re}_p{\bm f}_0$ in  (\ref{distflow}), where  $\bm{f}_0=\boldsymbol{w}^{(0)} \cdot \nabla \boldsymbol{w}^{(1)}_0+\boldsymbol{w}^{(1)}_0\cdot \nabla \boldsymbol{w}^{(0)}+\boldsymbol{w}^{(1)}_0\cdot \nabla\boldsymbol{w}^{(1)}_0$ is the leading-order nonlinear forcing of the disturbance flow. Note also that  $w_0^{(1)}$ is purely oscillatory, while $w_1^{(1)}$ has a non-zero time-average, exemplified by the flow in Fig.~\ref{figure1}c (bottom).
	
	Forces on the particle, as integrals of the fluid stress tensor over the particle surface $S_p$, are also expanded in this fashion. Application of a reciprocal theorem \cite{lovalenti1993hydrodynamic} formally yields the inertial force components as volume integrals over the entire fluid domain without the need to explicitly compute the flow field at that order. The reciprocal theorem employs a known test flow $\bm{u}'=u'(t)\bm{e}$ in a chosen direction $\bm{e}$. The component of the equation of particle motion in that direction, to $\O (\operatorname{Re}_p)$, is then
	\begin{subequations}
		\begin{align}
			m_p \frac{d U_p}{dt}&\!=\!F^{(0)}_0 \!+\!F^{(1)}_0\!+\!\operatorname{Re}_p (F^{(0)}_1\!+\!F^{(1)}_1)+\mathcal{O}(\operatorname{Re}_p^2),\\
			F^{(0)}_0&= \frac{F_S}{6\pi} \int_{V} \left( 3\lambda \partial_t \bar{\bm{u}}\right)\cdot \bm{e} dV,\label{F00}\\
			F^{(1)}_0&=\frac{F_S}{6\pi} \mathcal{L}^{-1}\left\{\int_{S_p}\frac{\left(\hat{\bm{u}}_{p}-\hat{\bar{\bm{u}}}\right)}{\hat{u}'}\cdot  ( \hat{\bm{\sigma}}'\cdot \bm{n})dS\right\},\label{F10}\\
			F^{(0)}_1&= \frac{F_S}{6\pi} \int_{V} \left(\bar{\bm{u}}\cdot \nabla \bar{\bm{u}}\right)\cdot \bm{e} dV,\label{F01}\\
			F^{(1)}_1&= - \frac{F_S}{6\pi}\mathcal{L}^{-1}\left\{ \frac{1}{\hat{u}'}\int_V \hat{\bm{u}}' \cdot \hat{\bm{f}}_0dV\right\},\label{F11}
		\end{align}\label{eom}\noindent
	\end{subequations}
	where $\bm{\sigma}'$ is the stress tensor of the test flow, hats denote Laplace transforms, and $\mathcal{L}^{-1}$ their inverse.
	All dimensional forces have the common Stokes drag scale  $F_S/6\pi=\nu \rho a_p U^{*}$.
	Equations \eqref{F00} and \eqref{F01} are forces exerted by the background flow, while \eqref{F10} and \eqref{F11} stem from the disturbance flow. The original MR equation contains $F^{(0)}_0$ and $F^{(1)}_0$, but only part of $F^{(0)}_1$, while $F^{(1)}_1$ is an entirely new term due to particle inertia. We shall show that these unrecognized contributions are not small corrections, but are dominant in relevant applications, particularly the inertial disturbance force $F^{(1)}_1$.

	This formalism is entirely general for arbitrary given background flows and provides (see Methods) analytical expressions for the new forces $F^{(0)}_1$ and $F^{(1)}_1$. The former reads
	\begin{align}
		\frac{F_{1}^{(0)}}{F_S}= \frac{4 }{9}\left(\bm{E}:\bm{G}\right)\cdot\bm{e}\,\mathcal{F}_1^{(0)}\,,\label{F01explicit}
	\end{align}
	where $\mathcal{F}_1^{(0)}=1/5$ \cite{ral20_MRNote_preprint}.
	The force $F^{(1)}_1$ simplifies considerably if the background flow around the particle is potential (this is fulfilled in almost all cases, requiring only that the distance between the particle and object surfaces is greater than the Stokes boundary layer thickness). If furthermore the particle is neutrally buoyant, we additionally obtain
	\begin{align}
		\frac{F_{1}^{(1)}}{F_S}= \frac{4 }{9}\left(\bm{E}:\bm{G}\right)\cdot\bm{e}\,\mathcal{F}_1^{(1)}(\lambda)\,,\label{F11Rep}
	\end{align}
	where the function
	$\mathcal{F}_1^{(1)} (\lambda)$ is determined analytically (see SI for details) and is universal, i.e., valid for arbitrary flow fields. While both \eqref{F01explicit} and \eqref{F11Rep} need non-zero gradient and curvature terms of the background flow,
	$\mathcal{F}_1^{(1)} (\lambda)$  captures
	the nonlinear effect of inertia of the leading order unsteady disturbance flow $\boldsymbol{w}^{(1)}_0$ on the particle. For micron-size particles where $\lambda\sim1$,  $\mathcal{F}_1^{(1)}$ is considerably larger than $\mathcal{F}_1^{(0)}$, so that \eqref{F11Rep} is the dominant effect in practical microfluidic applications.
	Adding both contributions \eqref{F01explicit} and \eqref{F11Rep}, the resulting dimensional force is \eqref{dimforce} before time-averaging.

	%
	
	\begin{figure*}[t]
		\centering
		\includegraphics[width=\textwidth]{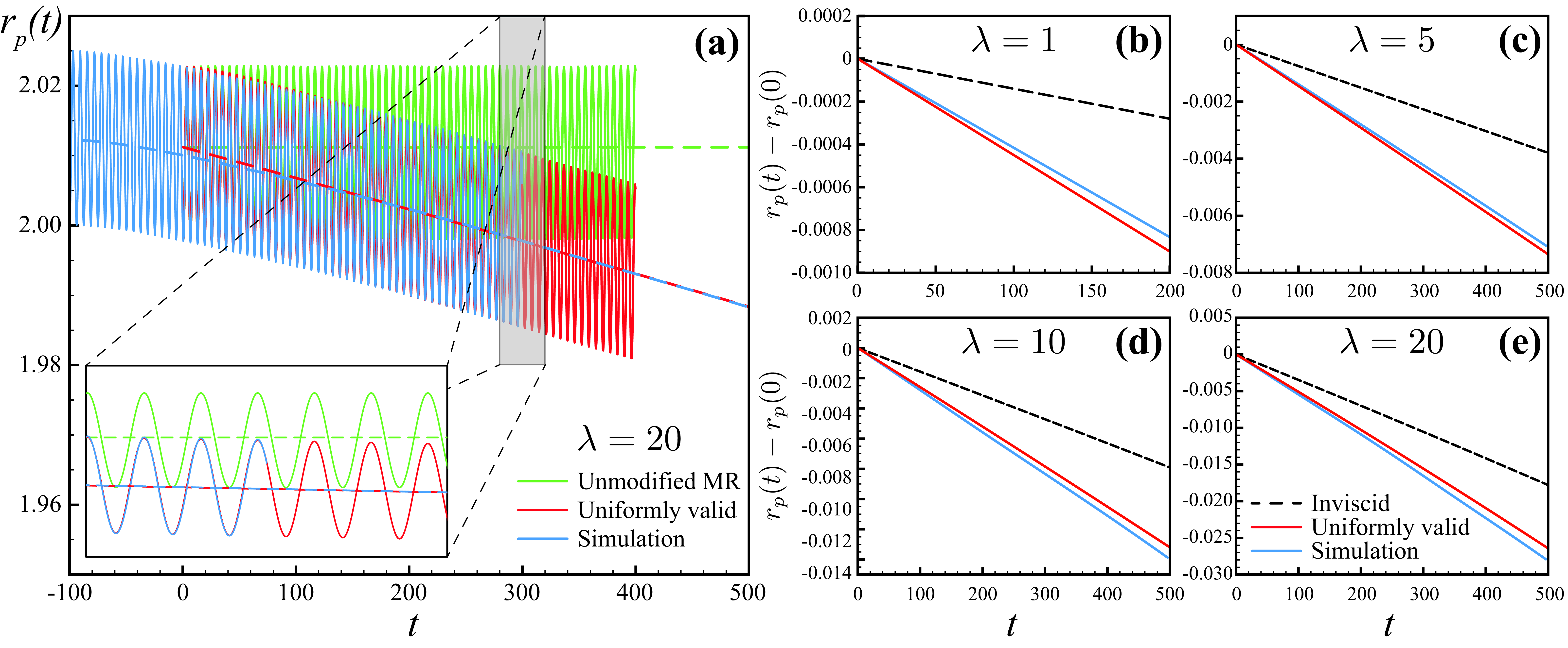}
		\caption{Comparison of theoretical (red) and simulated (blue) particle dynamics (radial displacements). (a) Full unsteady dynamics (solid lines) from DNS and theory Eq.\ \eqref{MRnondim} and time-averaged dynamics (dashed lines; theory uses Eq.\ \eqref{eom_slowtime} with \eqref{twotermforce}). The classical MR equation solutions (green) fail to even qualitatively capture the particle attraction to the bubble.
			(b-e) Steady dynamics from the uniformly valid asymptotic theory agrees with DNS for the entire range of $\lambda$ values. Dashed lines show the inviscid-limit theory, demonstrating significant quantitative discrepancies even for the largest $\lambda$.} \label{figure3}
	\end{figure*}
	
	
	
	
	We now turn to the prototypical oscillatory flow example of Fig.~\ref{figure1}c. This flow field's unique scale is the bubble radius ($L_\Gamma=L_\kappa=a_b$). With an oscillation amplitude of $\epsilon a_b$ ($\epsilon\ll 1$ in practical situations) the velocity scale is $U^*=\epsilon a_b \omega$,  
	and we anticipate that the relevant rectified (time-averaged) force will be proportional to $\epsilon^2$ (cf.\ \cite{agarwal2018inertial}). It is advantageous to change the length scale to $a_b$ here, introducing $\alpha\equiv a_p/a_b$, and to change the coordinate origin to the bubble center, so that the background flow has only one component $\bar{u}=\sin t/r^2$ in the direction ${\bm e}={\bm e}_r$. The oscillatory forces and the particle motion now follow explicitly (see Methods).

	

	Our ultimate goal is to predict the rectified trajectory of the particle after time-averaging over the fast oscillatory time scale, to provide practically useful guidance for precision applications.
	Time scale separation using the \emph{slow time}  $T=\epsilon^2 t$ analogous to \cite{agarwal2018inertial} (see Methods) obtains the leading order equation for rectified particle motion $r_{p}(T)$
	\begin{align}
		\frac{d r_{p}}{dT} = -\frac{6}{r_{p}^7}\alpha^2 \lambda \mathcal{F}(\lambda)\,,   \label{eom_slowtime}
	\end{align}
	where $\mathcal{F}(\lambda)=\mathcal{F}_1^{(1)}(\lambda)+\mathcal{F}^{(0)}_1$; \eqref{eom_slowtime} is readily solved analytically and is analogous to the result \eqref{particleeom}.
	Indeed, while the analytical form of the universal function $\mathcal{F}_1^{(1)}$ is complicated (see SI), one can Taylor expand in both the viscous limit ($\lambda \to 0$) and the inviscid limit ($\lambda \to \infty$) to obtain
	\begin{align}
		\mathcal{F}^{v} = \frac{9}{16}\sqrt{\frac{3}{2\lambda}}+\mathcal{O}(1),\quad \mathcal{F}^{i}=\frac{1}{3} + \mathcal{O}(1/\sqrt{\lambda}).
		\label{Flamexp}
	\end{align}
	Simply adding leading terms yields the uniformly valid expression (\ref{twotermforce}) for the total dimensionless force $\mathcal{F}(\lambda)$ on the particle. Note that our derivation is based fundamentally on the presence of both viscous and inertial effects, so that even $\mathcal{F}^{v}$ is a finite-inertia force. Its $\lambda^{-1/2}$ scaling for small $\lambda$ is analogous to Saffman's lift force \cite{saffman1965lift}, but is obtained without decomposing the domain into viscous and inertial regions (see Methods). Remarkably, the opposite limit $\mathcal{F}^{i}$ exactly asymptotes to the result obtained from the purely inviscid formalism of \cite{agarwal2018inertial} as $\lambda\to\infty$.

	We now demonstrate that \eqref{twotermforce} is accurate over the entire range of Stokes numbers by comparing our theory with independent, large-scale, 3D numerical simulations, previously validated in a range of streaming scenarios \cite{parthasarathy2019streaming,bhosale_parthasarathy_gazzola_2020}.  Fig.~\ref{figure2}a-e illustrate the rich time-averaged flow $\langle \boldsymbol{w}\rangle$ at different $\lambda$, while Fig.~\ref{figure2}g,i exemplify the expected confinement of vorticity around the particle. The simulations also serve to justify our omission of an inertia-dominated outer region (Fig.~\ref{figure2}f,h). In Fig.~\ref{figure3}a, we compare analytical and simulated particle trajectories on both the oscillatory and slow time scales. The classical MR equation fails to capture any of the attraction observed in  DNS, while the present theory is in excellent agreement both for the instantaneous motion and the rectified drift of the particle. Moreover, it succeeds over the entire range of $\lambda$ values, cf.\ Fig.~\ref{figure3}b-e. We see here that
	the inviscid formalism of Ref.~\cite{agarwal2018inertial}  (dashed lines) gives a much too weak attraction, particularly for practically relevant $\lambda\sim 1$.
	This is an intuitive outcome of taking viscosity into account, as the Stokes boundary layer (cf.\ Fig.~\ref{figure2}g,i) effectively increases particle size, so that forces scaling with particle size (cf.\ \eqref{dimforce}) become larger.
	Figure~\ref{figure3} also illustrates the great benefit of the analytical theory \eqref{eom_slowtime}, as individual DNS incur massive computational cost up to $\sim 100,\!000$ core-hours on the Stampede2 supercomputer (see SI).

\begin{figure}[hbt]
		\centering
		\includegraphics[width=0.48\textwidth]{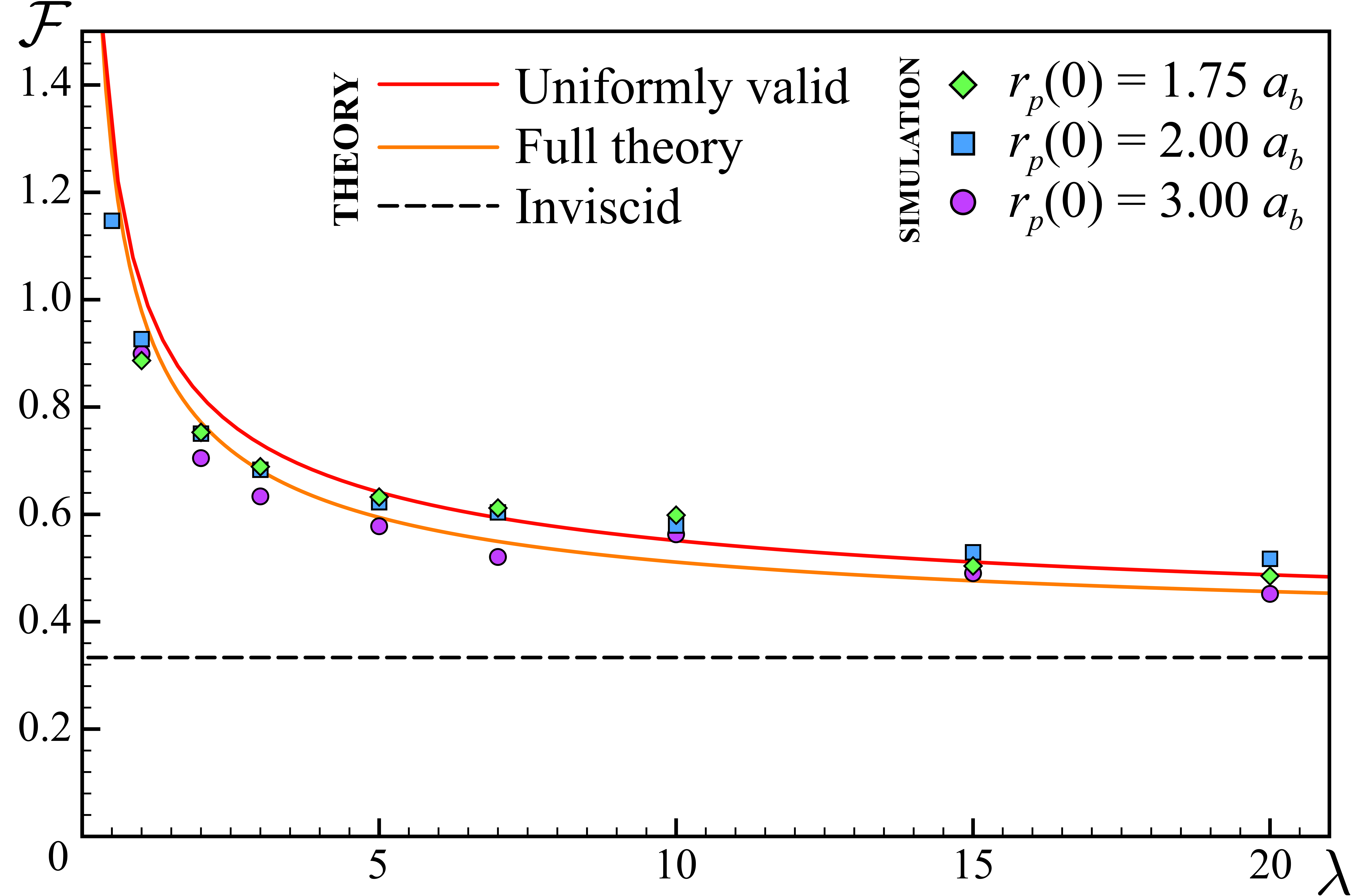}
		\caption{Comparison of the overall inertial force magnitude ${\mathcal F}$ in theory (lines) and simulation (symbols), for various $\lambda$ and initial particle positions $r_p(0)$. The uniformly valid expression (red) is extremely close to the full solution (orange) and in excellent agreement with all DNS data, while the inviscid theory (black dashed) severely underestimates the forces.} \label{figure4}
	\end{figure}

	Figure~\ref{figure4} summarizes the comparison between theory and  simulations: Time-averaged DNS trajectories (beyond an initial transient  -- see SI for details) for different values of $\lambda$ were fitted to (\ref{eom_slowtime}) to determine the dimensionless force
	${\mathcal F}$. Our analytical predictions are in quantitative agreement with DNS across the range of $\lambda$, exhibiting an average error of $\approx 7\%$.

	Finally, we emphasize that the novel forcing terms  \eqref{F01explicit} and \eqref{F11Rep}, investigated here in isolation, are not small corrections relative to the original MR terms in many commonplace applications. Indeed, for customary microfluidic settings, they are an order of magnitude stronger than density contrast induced forces, acoustofluidic radiation forces, or Fax\'en forces -- see the Methods section for a quantitative analysis.
	
	In summary, motivated by advancements in microfluidics, the present work reveals previously unrecognized forces acting on particles in viscous flows with finite inertial effects from oscillatory driving. These forces stem from flow gradients and curvatures, are attractive towards the oscillating object under mild assumptions, are much stronger than inviscid forces, and can be dominant over classically understood MR terms. They lead to significant displacements of cell-sized particles ($1-10\mu m$) over ms time scales, making them a promising tool for precision manipulation strategies. Our analysis shows that a surprisingly simple expression accurately predicts particle motion, as quantitatively confirmed against first-principle, large-scale direct numerical simulations. The theory highlights the immense reduction in computational effort between DNS and an explicit analytical theory, and as a generalization of the Maxey-Riley formalism is applicable to a wide variety of flow situations.
	
	\newpage
	
	{\bf Acknowledgments:} The authors thank Kaitlyn Hood, Gabriel Juarez, and Howard A.\ Stone for fruitful discussions. The authors also acknowledge support by the National Science Foundation under NSF CAREER Grant No. CBET-1846752 (MG) and by the Blue Waters project (OCI-0725070, ACI-1238993), a joint effort of the University of Illinois at Urbana-Champaign and its National Center for Supercomputing Applications.
	This work also used the Extreme Science and Engineering Discovery Environment (XSEDE) \citep{Towns:2014} Stampede2, supported by National Science Foundation grant no. ACI-1548562, at the Texas Advanced Computing Center (TACC) through allocation TG-MCB190004.
	
	\vspace*{5mm}
	
	
	\vspace*{5mm}
	
	\centerline{\bf Methods}
	
	\small
	\vspace*{5mm}
	
	{\bf General solutions and the Reciprocal Theorem}.
	The leading-order oscillatory disturbance flow field $\boldsymbol{w}^{(1)}_0$ is obtained by inserting \eqref{ubarexp} into the leading order of \eqref{distflow} and can be formally expressed as a series solution \cite{landau1959course,pozrikidis1992boundary}
	\begin{align}
		\boldsymbol{w}^{(1)}_0 = \bm{\mathcal{M}}_D \cdot \bm{u}_s  - \bm{\mathcal{M}}_Q \cdot\left(\bm{r}\cdot \bm{E}\right) -\bm{\mathcal{M}}_O \cdot\left(\bm{r}\bm{r}:\bm{G}\right) + \dots, \label{w10gen}
	\end{align}
	where $\bm{u}_s=\bm{u}_{p_0} -\bm{\bar{u}}|_{\bm{r}_{p_0}}$ is the slip velocity and $\bm{\mathcal{M}}_{D,Q,O}(\bm{r},\lambda)$ are spatially dependent mobility tensors independent of the particular background flow -- see SI for explicit expressions in the case of harmonic oscillatory flows. All information about the specific background flow is contained in the constant quantities $\bm{u}_s$, $\bm{E}$, and $\bm{G}$. The $\mathcal{O}(\operatorname{Re}_p)$ flow field  $\boldsymbol{w}^{(1)}_1$ does not need to be computed explicitly; instead, we use a reciprocal theorem. Denoting Laplace-transformed quantities by hats, application of the divergence theorem results in the following symmetry relation:
	\begin{align}
		&\oint_S  (\hat{\boldsymbol{w}}^{(1)}_1\cdot \hat{\bm{\sigma}}' - \hat{\bm{u}}'\cdot \hat{\bm{\sigma}}^{(1)}_1)\cdot \bm{m} \, dS\nonumber\\
		&= \int_V \left[\nabla \cdot (\hat{\boldsymbol{w}}^{(1)}_1\cdot \hat{\bm{\sigma}}') - \nabla \cdot (\hat{\bm{u}}'\cdot \hat{\bm{\sigma}}^{(1)}_1)\right]dV.
	\end{align}
	As shown in the SI, the above expression yields the $\mathcal{O}(\operatorname{Re}_p)$ force on the particle captured by \eqref{F11}. We note that the computation of the
	volume integral simplifies considerably: the integrand is proportional to $\bm{f}_0$, in which
	only certain products are non-vanishing when the angular integration around the particle is performed. For instance, the first term in $\bm{f}_0$ is
	$\left(\boldsymbol{\bar{u}}-\bm{u}_{p_0}\right) \cdot \nabla \boldsymbol{w}^{(1)}_0 = \left(-\bm{u}_s  + \bm{r}\cdot \bm{E}+ \bm{r}\bm{r}:\bm{G} \right)\cdot \nabla\left(\bm{\mathcal{M}}_D \cdot \bm{u}_s- \bm{\mathcal{M}}_Q \cdot\left(\bm{r}\cdot \bm{E}\right) -\bm{\mathcal{M}}_O \cdot\left(\bm{r}\bm{r}:\bm{G}\right) \right)$. Due to the alternating symmetry of terms in the background flow and consequently $\boldsymbol{w}^{(1)}_0$, only products of adjacent terms survive integration, while e.g.\ a term involving $\bm{u}_s\cdot \nabla \left(\bm{\mathcal{M}}_D \cdot\bm{u}_s\right)$ vanishes after volume integration.
	
	\vspace*{5mm}

	{\bf Simplification for neutrally buoyant particles}.
	In this work, we restrict our analysis to the case of neutrally buoyant particles. Consequently, the slip velocity $\bm{u}_s$ vanishes, so that only tensor products involving $\bm{E}$ and $\bm{G}$ contribute to the volume integral (given in SI). Furthermore, in typical microfluidic applications, the particle experiences an oscillatory background flow that is potential: to excellent approximation, this holds true if the particle is outside the Stokes boundary layer of the oscillating object, that is if $h_p=r_p-a_p-a_b \gtrsim \delta_S$, which simplifies to the easily satisfied condition $\lambda\gtrsim (a_p/h_p)^2$.  Then, the only way to construct a force vector from a contraction of the higher rank tensors $\bm{E}$ and $\bm{G}$ in a potential flow is $\bm{E}:\bm{G}$ (cf. \cite{danilov2000mean,nadim1991motion}), which can be pulled out of the integral as a common factor. Thus, the volume integral can be evaluated generally for {\em any} background flow field and results in the single term \eqref{F11Rep} with the universal function $\mathcal{F}_1^{(1)}(\lambda)$.
	
	\vspace*{5mm}
	
	{\bf Inner-Outer (wake) formalism}.
	Often, the evaluation of forces on particles in a flow is complicated by the transition between a viscous-dominated inner flow volume (near the particle) and an inertia-dominated outer volume, necessitating an asymptotic matching of the two limits (such as for the Oseen \cite{lovalenti1993hydrodynamic} and Saffman \cite{saffman1965lift} problems). The present formalism, however, only employs an inner-solution expansion and still obtains highly accurate predictions (see also \cite{hood2015inertial}). This behavior can be rationalized by invoking the analysis of Lovalenti and Brady \cite{lovalenti1993hydrodynamic} who showed that an outer region does not occur when the characteristic unsteady time scale $\omega^{-1}$ is shorter than the convective inertial time scale $\nu/(U^*w^{(0)})^2$, where $w^{(0)}$ is the dimensionless velocity scale of the fluid as measured in the particle reference frame. For density matched particles $w^{(0)} = {\cal O}(\alpha)$, so that this criterion reduces to $\epsilon^2 \lambda \ll 1$, requiring the oscillation amplitude of the flow to be smaller than $\delta_S \alpha^{-1}$, which is easily satisfied in most experimental situations. More directly, the Lovalenti-Brady criterion relies on the magnitude of oscillatory inertia in the disturbance flow $\partial \boldsymbol{w}^{(1)}/\partial t$ being much larger than that of the advective term ${\bm f}$. DNS verifies that this relation holds for the entire range of $\lambda$ treated here (see Fig.~\ref{figure2}f,h). As a separate effect, outer flow inertia due to the slow (steady) motion of the particle will be present, but only results in $O(\epsilon)$ corrections to the Stokes drag.
	
	\vspace*{5mm}

	{\bf Oscillatory equation of motion in radial flow}. For the special case of the bubble executing pure breathing oscillations with the radial flow field $\bar{u}=\sin t/r^2$, it is straightforward to compute $\bm{E}:\bm{G}\cdot\bm{e}_r = -18\sin^2t/r_{p}^7$, where $r_{p}(t)$ is the instantaneous particle position.
	Using (\ref{eom}), (\ref{F01explicit}), (\ref{F11Rep}), and noting $\alpha\operatorname{Re}_p=3\epsilon\lambda$, the non-dimensional equation of motion for $r_p(t)$ of a neutrally buoyant particle explicitly reads:
	\begin{align}
		&\lambda \frac{d^2 r_p}{d t^2} =\epsilon \lambda\left(\frac{\cos t}{r_{p}^2} - 2\epsilon\frac{\sin^2 t}{r_{p}^5}\right)-\frac{2\lambda}{3}\epsilon^2 \alpha^2 \frac{18\sin^2 t}{r_{p}^7}\mathcal{F}^{(0)}\nonumber\\
		&+ \left[\frac{\sin t}{r_{p}^2}-\frac{d r_p}{d t}\right] -\left[\frac{2\lambda}{3}\epsilon^2 \alpha^2 \frac{(18\sin^2 t)}{r_{p}^7}\mathcal{F}_1^{(1)}\lambda)\right]\,,  \label{MRnondim}
	\end{align}
	where the first line on the RHS represents contributions from $F_0^{(0)}$ and $F_1^{(0)}$, while the first and second terms in square parentheses represent $F_0^{(1)}$  and $F_1^{(1)}$, respectively. Note that, for neutrally buoyant particles, the time-periodic character of the flow precludes memory terms that would otherwise emerge from the inverse Laplace transforms \cite{basset1888treatise,maxey1983equation,lovalenti1993hydrodynamic}.
	
	\vspace*{5mm}

	{\bf Time scale separation and time averaging}.
	Assuming  $\epsilon\ll 1$, we introduce the slow time $T=\epsilon^2 t$, in addition to the fast time $t$. Using the following transformations
	\begin{subequations}
		\begin{alignat}{4}
			&r_p(t) \mapsto r_p(t,T),\\
			&\frac{d}{d t} \mapsto \frac{\partial }{\partial t} +\epsilon^2\frac{\partial }{\partial T},\\
			&\frac{d^2}{d t^2} \mapsto \frac{\partial^2 }{\partial t^2} +2\epsilon^2\frac{\partial^2 }{\partial t \partial T} +\epsilon^4\frac{\partial^2 }{\partial T^2},
		\end{alignat}
	\end{subequations}
	we seek a perturbation solution in $\epsilon$ of the general form $r_p(t,T)=r_{p}(T)+\epsilon \check{r}_{p}(t,T)+\epsilon^2 \check{\check{r}}_{p}(t,T)+\dots $, and separate orders in \eqref{MRnondim}. The procedure is outlined in \cite{agarwal2018inertial} and results in a leading-order equation for $r_{p}(T)$ given by \eqref{eom_slowtime}, dependent on the slow time scale only (the scale t being averaged out).
	
	\vspace*{5mm}

	{\bf Simulation method and numerical implementation}.
	Here we briefly describe the governing equations and numerical technique used in our simulations.
	We consider two spherical bodies (an oscillating microbubble and a neutrally buoyant particle) immersed in an unbounded domain of incompressible viscous fluid.
	We denote the computational domain as $\Omega = \Omega_\text{f} \cup \Omega_\text{B}$, where $\Omega_\text{f}$ is the fluid domain and $\Omega_\text{B} = \Omega_\text{b} \cup \Omega_\text{p}$ is the domain in which the bubble ($\Omega_\text{b}$) and particle ($\Omega_\text{p}$) reside, and denote the interface between the fluid and the bodies as $\partial \Omega_\text{B}$.
	The flow is then described by the incompressible Navier--Stokes equation
	\begin{equation}
		\bnabla \bcdot \bm{u} = 0, \quad
		\frac{\partial \bm{u}}{\partial t} + \left( \bm{u} \bcdot \bnabla \right)\bm{u} = -\frac{1}{\rho}\nabla p + \nu \bnabla^2 \bm{u}~~~~\bm{x}\in\Omega\setminus\Omega_\text{B}
		\label{eq:navier-stokes}
	\end{equation}
	where $\rho$, $p$, $\bm{u}$ and $\nu$ are the fluid density, pressure, velocity and kinematic viscosity, respectively. We impose the no-slip boundary condition $\bm{u} = \bm{u}_B$ at $\partial \Omega_B$, where $\bm{u}_B$ is the body velocity, and feedback from the fluid to the body is described by Newton's equation of motion. The system of equations is solved in velocity--vorticity form using the remeshed vortex method combined with Brinkmann penalization and a projection approach \cite{gazzola2011simulations}. This method has been extensively validated across a range of fluid--structure interaction problems, from flow past bluff bodies to biological swimming \cite{gazzola2011simulations,gazzola2012flow,gazzola2012c,gazzola2014reinforcement,gazzola2016learning}. Recently, the accuracy of this method has been demonstrated in rectified flow contexts as well, capturing steady streaming responses from arbitrary shapes in 2D and 3D \cite{bhosale_parthasarathy_gazzola_2020,parthasarathy2019streaming}. More details on method implementation and simulation techniques can be found in the SI.
	
	\vspace*{5mm}

	{\bf Comparison with other hydrodynamic forces}.
	We have discussed a special case of radial symmetry quantitatively because it isolates the novel inertial forces reported here as the only effect, thus allowing us to assess the accuracy of the theory. In more general flow situations, other forces will compete with $F_{\Gamma\kappa}$, and we assess their relative magnitude here. If the particle density $\rho_p$ does not match $\rho$, a density contrast force \cite{agarwal2018inertial} is induced, generalizing acoustofluidic secondary radiation forces. In order for this force to exceed $F_{\Gamma\kappa}$, the density contrast needs to fulfill $\rho_p/\rho - 1\gtrsim 3(a_p/r_p)^2(1+2/\sqrt{\lambda})$. Appreciable forces only act when $r_p\gtrsim a_b$ and if $\lambda$ is not very small; thus, $\rho_p/\rho - 1 \gtrsim 0.3$ for typical geometries characterized by $\alpha\lesssim 0.2$. In most microfluidic, and certainly in biomedical applications, the density contrast is far less: even at $5\%$ density difference (e.g.\ for polystyrene particles), $F_{\Gamma\kappa}$ is 5-30 times stronger than the density contrast force for $0.5 < \lambda < 5$. Other forces result from steady flows: oscillation of an $a_b$-sized object will generically induce steady streaming flow at speed $\sim \epsilon^2 a_b U^*$, and it may have transverse gradients of scale $a_b$ (in addition to radial gradients). This situation induces a Saffman lift force $L_S$ \cite{saffman1965lift} for particles with finite slip velocity $V_s$ (again because of density mismatch) \cite{lovalenti1993hydrodynamic,agarwal2018inertial}. $L_S$ and $F_{\Gamma\kappa}$ are of equal magnitude if $V_s\sim 5\alpha^2(4.1+2\sqrt{\lambda})U^{*}$. In realistic settings, $V_s$ would need to exceed $U^*$, implying that the steady flow would overwhelm the oscillatory motion, defeating the purpose of oscillatory-flow microfluidics. Lastly, flows with finite $ \nabla^2{\bf \bar{U}}$ give rise to Fax\'en terms in added mass and drag. However, the oscillatory  flows discussed here are (almost) potential flows as shown above, so that the leading order effect of Fax\'en terms comes from steady flow curvature and provides only an ${\cal O}(\alpha^2)$ correction to the steady-flow Stokes drag.
	We conclude that the inertial force described here is the dominant effect in many realistic oscillating microfluidics applications.

	\bibliographystyle{apsrev4-2}
	
	\bibliography{ms}

\providecommand{\noopsort}[1]{}\providecommand{\singleletter}[1]{#1}%
\begin{thebibliography}{43}%
\makeatletter
\providecommand \@ifxundefined [1]{%
 \@ifx{#1\undefined}
}%
\providecommand \@ifnum [1]{%
 \ifnum #1\expandafter \@firstoftwo
 \else \expandafter \@secondoftwo
 \fi
}%
\providecommand \@ifx [1]{%
 \ifx #1\expandafter \@firstoftwo
 \else \expandafter \@secondoftwo
 \fi
}%
\providecommand \natexlab [1]{#1}%
\providecommand \enquote  [1]{``#1''}%
\providecommand \bibnamefont  [1]{#1}%
\providecommand \bibfnamefont [1]{#1}%
\providecommand \citenamefont [1]{#1}%
\providecommand \href@noop [0]{\@secondoftwo}%
\providecommand \href [0]{\begingroup \@sanitize@url \@href}%
\providecommand \@href[1]{\@@startlink{#1}\@@href}%
\providecommand \@@href[1]{\endgroup#1\@@endlink}%
\providecommand \@sanitize@url [0]{\catcode `\\12\catcode `\$12\catcode
  `\&12\catcode `\#12\catcode `\^12\catcode `\_12\catcode `\%12\relax}%
\providecommand \@@startlink[1]{}%
\providecommand \@@endlink[0]{}%
\providecommand \url  [0]{\begingroup\@sanitize@url \@url }%
\providecommand \@url [1]{\endgroup\@href {#1}{\urlprefix }}%
\providecommand \urlprefix  [0]{URL }%
\providecommand \Eprint [0]{\href }%
\providecommand \doibase [0]{https://doi.org/}%
\providecommand \selectlanguage [0]{\@gobble}%
\providecommand \bibinfo  [0]{\@secondoftwo}%
\providecommand \bibfield  [0]{\@secondoftwo}%
\providecommand \translation [1]{[#1]}%
\providecommand \BibitemOpen [0]{}%
\providecommand \bibitemStop [0]{}%
\providecommand \bibitemNoStop [0]{.\EOS\space}%
\providecommand \EOS [0]{\spacefactor3000\relax}%
\providecommand \BibitemShut  [1]{\csname bibitem#1\endcsname}%
\let\auto@bib@innerbib\@empty
\bibitem [{\citenamefont {Ho}\ and\ \citenamefont
  {Leal}(1974)}]{ho1974inertial}%
  \BibitemOpen
  \bibfield  {author} {\bibinfo {author} {\bibfnamefont {B.}~\bibnamefont
  {Ho}}\ and\ \bibinfo {author} {\bibfnamefont {L.}~\bibnamefont {Leal}},\
  }\href@noop {} {\bibfield  {journal} {\bibinfo  {journal} {Journal of fluid
  mechanics}\ }\textbf {\bibinfo {volume} {65}},\ \bibinfo {pages} {365}
  (\bibinfo {year} {1974})}\BibitemShut {NoStop}%
\bibitem [{\citenamefont {Lovalenti}\ and\ \citenamefont
  {Brady}(1993)}]{lovalenti1993hydrodynamic}%
  \BibitemOpen
  \bibfield  {author} {\bibinfo {author} {\bibfnamefont {P.~M.}\ \bibnamefont
  {Lovalenti}}\ and\ \bibinfo {author} {\bibfnamefont {J.~F.}\ \bibnamefont
  {Brady}},\ }\href@noop {} {\bibfield  {journal} {\bibinfo  {journal} {Journal
  of Fluid Mechanics}\ }\textbf {\bibinfo {volume} {256}},\ \bibinfo {pages}
  {561} (\bibinfo {year} {1993})}\BibitemShut {NoStop}%
\bibitem [{\citenamefont {Schonberg}\ and\ \citenamefont
  {Hinch}(1989)}]{schonberg1989inertial}%
  \BibitemOpen
  \bibfield  {author} {\bibinfo {author} {\bibfnamefont {J.~A.}\ \bibnamefont
  {Schonberg}}\ and\ \bibinfo {author} {\bibfnamefont {E.}~\bibnamefont
  {Hinch}},\ }\href@noop {} {\bibfield  {journal} {\bibinfo  {journal} {Journal
  of Fluid Mechanics}\ }\textbf {\bibinfo {volume} {203}},\ \bibinfo {pages}
  {517} (\bibinfo {year} {1989})}\BibitemShut {NoStop}%
\bibitem [{\citenamefont {Di~Carlo}(2009)}]{di2009inertial}%
  \BibitemOpen
  \bibfield  {author} {\bibinfo {author} {\bibfnamefont {D.}~\bibnamefont
  {Di~Carlo}},\ }\href@noop {} {\bibfield  {journal} {\bibinfo  {journal} {Lab
  on a Chip}\ }\textbf {\bibinfo {volume} {9}},\ \bibinfo {pages} {3038}
  (\bibinfo {year} {2009})}\BibitemShut {NoStop}%
\bibitem [{\citenamefont {Hood}\ \emph {et~al.}(2015)\citenamefont {Hood},
  \citenamefont {Lee},\ and\ \citenamefont {Roper}}]{hood2015inertial}%
  \BibitemOpen
  \bibfield  {author} {\bibinfo {author} {\bibfnamefont {K.}~\bibnamefont
  {Hood}}, \bibinfo {author} {\bibfnamefont {S.}~\bibnamefont {Lee}},\ and\
  \bibinfo {author} {\bibfnamefont {M.}~\bibnamefont {Roper}},\ }\href@noop {}
  {\bibfield  {journal} {\bibinfo  {journal} {Journal of Fluid Mechanics}\
  }\textbf {\bibinfo {volume} {765}},\ \bibinfo {pages} {452} (\bibinfo {year}
  {2015})}\BibitemShut {NoStop}%
\bibitem [{\citenamefont {Lutz}\ \emph {et~al.}(2006)\citenamefont {Lutz},
  \citenamefont {Chen},\ and\ \citenamefont {Schwartz}}]{lutz2006hydrodynamic}%
  \BibitemOpen
  \bibfield  {author} {\bibinfo {author} {\bibfnamefont {B.~R.}\ \bibnamefont
  {Lutz}}, \bibinfo {author} {\bibfnamefont {J.}~\bibnamefont {Chen}},\ and\
  \bibinfo {author} {\bibfnamefont {D.~T.}\ \bibnamefont {Schwartz}},\
  }\href@noop {} {\bibfield  {journal} {\bibinfo  {journal} {Analytical
  chemistry}\ }\textbf {\bibinfo {volume} {78}},\ \bibinfo {pages} {5429}
  (\bibinfo {year} {2006})}\BibitemShut {NoStop}%
\bibitem [{\citenamefont {Rogers}\ and\ \citenamefont
  {Neild}(2011)}]{rogers2011selective}%
  \BibitemOpen
  \bibfield  {author} {\bibinfo {author} {\bibfnamefont {P.}~\bibnamefont
  {Rogers}}\ and\ \bibinfo {author} {\bibfnamefont {A.}~\bibnamefont {Neild}},\
  }\href@noop {} {\bibfield  {journal} {\bibinfo  {journal} {Lab on a Chip}\
  }\textbf {\bibinfo {volume} {11}},\ \bibinfo {pages} {3710} (\bibinfo {year}
  {2011})}\BibitemShut {NoStop}%
\bibitem [{\citenamefont {Thameem}\ \emph {et~al.}(2017)\citenamefont
  {Thameem}, \citenamefont {Rallabandi},\ and\ \citenamefont
  {Hilgenfeldt}}]{thameem2017fast}%
  \BibitemOpen
  \bibfield  {author} {\bibinfo {author} {\bibfnamefont {R.}~\bibnamefont
  {Thameem}}, \bibinfo {author} {\bibfnamefont {B.}~\bibnamefont
  {Rallabandi}},\ and\ \bibinfo {author} {\bibfnamefont {S.}~\bibnamefont
  {Hilgenfeldt}},\ }\href@noop {} {\bibfield  {journal} {\bibinfo  {journal}
  {Physical Review Fluids}\ }\textbf {\bibinfo {volume} {2}},\ \bibinfo {pages}
  {052001} (\bibinfo {year} {2017})}\BibitemShut {NoStop}%
\bibitem [{\citenamefont {Maxey}\ and\ \citenamefont
  {Riley}(1983)}]{maxey1983equation}%
  \BibitemOpen
  \bibfield  {author} {\bibinfo {author} {\bibfnamefont {M.~R.}\ \bibnamefont
  {Maxey}}\ and\ \bibinfo {author} {\bibfnamefont {J.~J.}\ \bibnamefont
  {Riley}},\ }\href {https://doi.org/10.1063/1.864230} {\bibfield  {journal}
  {\bibinfo  {journal} {The Physics of Fluids}\ }\textbf {\bibinfo {volume}
  {26}},\ \bibinfo {pages} {883} (\bibinfo {year} {1983})}\BibitemShut
  {NoStop}%
\bibitem [{\citenamefont {Lutz}\ \emph {et~al.}(2005)\citenamefont {Lutz},
  \citenamefont {Chen},\ and\ \citenamefont {Schwartz}}]{lutz2005microscopic}%
  \BibitemOpen
  \bibfield  {author} {\bibinfo {author} {\bibfnamefont {B.~R.}\ \bibnamefont
  {Lutz}}, \bibinfo {author} {\bibfnamefont {J.}~\bibnamefont {Chen}},\ and\
  \bibinfo {author} {\bibfnamefont {D.~T.}\ \bibnamefont {Schwartz}},\
  }\href@noop {} {\bibfield  {journal} {\bibinfo  {journal} {Physics of
  Fluids}\ }\textbf {\bibinfo {volume} {17}},\ \bibinfo {pages} {023601}
  (\bibinfo {year} {2005})}\BibitemShut {NoStop}%
\bibitem [{\citenamefont {Wang}\ \emph {et~al.}(2011)\citenamefont {Wang},
  \citenamefont {Jalikop},\ and\ \citenamefont {Hilgenfeldt}}]{wang2011size}%
  \BibitemOpen
  \bibfield  {author} {\bibinfo {author} {\bibfnamefont {C.}~\bibnamefont
  {Wang}}, \bibinfo {author} {\bibfnamefont {S.~V.}\ \bibnamefont {Jalikop}},\
  and\ \bibinfo {author} {\bibfnamefont {S.}~\bibnamefont {Hilgenfeldt}},\
  }\href@noop {} {\bibfield  {journal} {\bibinfo  {journal} {Applied Physics
  Letters}\ }\textbf {\bibinfo {volume} {99}},\ \bibinfo {pages} {034101}
  (\bibinfo {year} {2011})}\BibitemShut {NoStop}%
\bibitem [{\citenamefont {Schmid}\ \emph {et~al.}(2014)\citenamefont {Schmid},
  \citenamefont {Weitz},\ and\ \citenamefont {Franke}}]{schmid2014sorting}%
  \BibitemOpen
  \bibfield  {author} {\bibinfo {author} {\bibfnamefont {L.}~\bibnamefont
  {Schmid}}, \bibinfo {author} {\bibfnamefont {D.~A.}\ \bibnamefont {Weitz}},\
  and\ \bibinfo {author} {\bibfnamefont {T.}~\bibnamefont {Franke}},\
  }\href@noop {} {\bibfield  {journal} {\bibinfo  {journal} {Lab on a Chip}\
  }\textbf {\bibinfo {volume} {14}},\ \bibinfo {pages} {3710} (\bibinfo {year}
  {2014})}\BibitemShut {NoStop}%
\bibitem [{\citenamefont {Chen}\ and\ \citenamefont
  {Lee}(2014)}]{chen2014manipulation}%
  \BibitemOpen
  \bibfield  {author} {\bibinfo {author} {\bibfnamefont {Y.}~\bibnamefont
  {Chen}}\ and\ \bibinfo {author} {\bibfnamefont {S.}~\bibnamefont {Lee}},\
  }\href@noop {} {\bibfield  {journal} {\bibinfo  {journal} {Integrative and
  comparative biology}\ }\textbf {\bibinfo {volume} {54}},\ \bibinfo {pages}
  {959} (\bibinfo {year} {2014})}\BibitemShut {NoStop}%
\bibitem [{\citenamefont {Park}\ \emph {et~al.}(2016)\citenamefont {Park},
  \citenamefont {Shin}, \citenamefont {Lee},\ and\ \citenamefont
  {Chung}}]{park2016chip}%
  \BibitemOpen
  \bibfield  {author} {\bibinfo {author} {\bibfnamefont {I.~S.}\ \bibnamefont
  {Park}}, \bibinfo {author} {\bibfnamefont {J.~H.}\ \bibnamefont {Shin}},
  \bibinfo {author} {\bibfnamefont {Y.~R.}\ \bibnamefont {Lee}},\ and\ \bibinfo
  {author} {\bibfnamefont {S.~K.}\ \bibnamefont {Chung}},\ }\href@noop {}
  {\bibfield  {journal} {\bibinfo  {journal} {Sensors and Actuators A:
  Physical}\ }\textbf {\bibinfo {volume} {248}},\ \bibinfo {pages} {214}
  (\bibinfo {year} {2016})}\BibitemShut {NoStop}%
\bibitem [{\citenamefont {Thameem}\ \emph {et~al.}(2016)\citenamefont
  {Thameem}, \citenamefont {Rallabandi},\ and\ \citenamefont
  {Hilgenfeldt}}]{thameem2016particle}%
  \BibitemOpen
  \bibfield  {author} {\bibinfo {author} {\bibfnamefont {R.}~\bibnamefont
  {Thameem}}, \bibinfo {author} {\bibfnamefont {B.}~\bibnamefont
  {Rallabandi}},\ and\ \bibinfo {author} {\bibfnamefont {S.}~\bibnamefont
  {Hilgenfeldt}},\ }\href@noop {} {\bibfield  {journal} {\bibinfo  {journal}
  {Biomicrofluidics}\ }\textbf {\bibinfo {volume} {10}},\ \bibinfo {pages}
  {014124} (\bibinfo {year} {2016})}\BibitemShut {NoStop}%
\bibitem [{\citenamefont {Volk}\ \emph {et~al.}(2020)\citenamefont {Volk},
  \citenamefont {Rossi}, \citenamefont {Rallabandi}, \citenamefont
  {K{\"a}hler}, \citenamefont {Hilgenfeldt},\ and\ \citenamefont
  {Marin}}]{volk2020size}%
  \BibitemOpen
  \bibfield  {author} {\bibinfo {author} {\bibfnamefont {A.}~\bibnamefont
  {Volk}}, \bibinfo {author} {\bibfnamefont {M.}~\bibnamefont {Rossi}},
  \bibinfo {author} {\bibfnamefont {B.}~\bibnamefont {Rallabandi}}, \bibinfo
  {author} {\bibfnamefont {C.~J.}\ \bibnamefont {K{\"a}hler}}, \bibinfo
  {author} {\bibfnamefont {S.}~\bibnamefont {Hilgenfeldt}},\ and\ \bibinfo
  {author} {\bibfnamefont {A.}~\bibnamefont {Marin}},\ }\href@noop {}
  {\bibfield  {journal} {\bibinfo  {journal} {Physical Review Fluids}\ }\textbf
  {\bibinfo {volume} {5}},\ \bibinfo {pages} {114201} (\bibinfo {year}
  {2020})}\BibitemShut {NoStop}%
\bibitem [{\citenamefont {Di~Carlo}\ \emph {et~al.}(2007)\citenamefont
  {Di~Carlo}, \citenamefont {Irimia}, \citenamefont {Tompkins},\ and\
  \citenamefont {Toner}}]{di2007continuous}%
  \BibitemOpen
  \bibfield  {author} {\bibinfo {author} {\bibfnamefont {D.}~\bibnamefont
  {Di~Carlo}}, \bibinfo {author} {\bibfnamefont {D.}~\bibnamefont {Irimia}},
  \bibinfo {author} {\bibfnamefont {R.~G.}\ \bibnamefont {Tompkins}},\ and\
  \bibinfo {author} {\bibfnamefont {M.}~\bibnamefont {Toner}},\ }\href@noop {}
  {\bibfield  {journal} {\bibinfo  {journal} {PNAS}\ }\textbf {\bibinfo
  {volume} {104}},\ \bibinfo {pages} {18892} (\bibinfo {year}
  {2007})}\BibitemShut {NoStop}%
\bibitem [{\citenamefont {Warkiani}\ \emph {et~al.}(2015)\citenamefont
  {Warkiani}, \citenamefont {Tay}, \citenamefont {Khoo}, \citenamefont
  {Xiaofeng}, \citenamefont {Han},\ and\ \citenamefont
  {Lim}}]{warkiani2015malaria}%
  \BibitemOpen
  \bibfield  {author} {\bibinfo {author} {\bibfnamefont {M.~E.}\ \bibnamefont
  {Warkiani}}, \bibinfo {author} {\bibfnamefont {A.~K.~P.}\ \bibnamefont
  {Tay}}, \bibinfo {author} {\bibfnamefont {B.~L.}\ \bibnamefont {Khoo}},
  \bibinfo {author} {\bibfnamefont {X.}~\bibnamefont {Xiaofeng}}, \bibinfo
  {author} {\bibfnamefont {J.}~\bibnamefont {Han}},\ and\ \bibinfo {author}
  {\bibfnamefont {C.~T.}\ \bibnamefont {Lim}},\ }\href@noop {} {\bibfield
  {journal} {\bibinfo  {journal} {Lab on a Chip}\ }\textbf {\bibinfo {volume}
  {15}},\ \bibinfo {pages} {1101} (\bibinfo {year} {2015})}\BibitemShut
  {NoStop}%
\bibitem [{\citenamefont {Bjerknes}(1906)}]{bjerknes1906fields}%
  \BibitemOpen
  \bibfield  {author} {\bibinfo {author} {\bibfnamefont {V.}~\bibnamefont
  {Bjerknes}},\ }\href@noop {} {\emph {\bibinfo {title} {Fields of force}}}\
  (\bibinfo  {publisher} {General Books},\ \bibinfo {year} {1906})\BibitemShut
  {NoStop}%
\bibitem [{\citenamefont {Hay}\ \emph {et~al.}(2009)\citenamefont {Hay},
  \citenamefont {Hamilton}, \citenamefont {Ilinskii},\ and\ \citenamefont
  {Zabolotskaya}}]{hay2009model}%
  \BibitemOpen
  \bibfield  {author} {\bibinfo {author} {\bibfnamefont {T.~A.}\ \bibnamefont
  {Hay}}, \bibinfo {author} {\bibfnamefont {M.~F.}\ \bibnamefont {Hamilton}},
  \bibinfo {author} {\bibfnamefont {Y.~A.}\ \bibnamefont {Ilinskii}},\ and\
  \bibinfo {author} {\bibfnamefont {E.~A.}\ \bibnamefont {Zabolotskaya}},\
  }\href@noop {} {\bibfield  {journal} {\bibinfo  {journal} {The Journal of the
  Acoustical Society of America}\ }\textbf {\bibinfo {volume} {125}},\ \bibinfo
  {pages} {1331} (\bibinfo {year} {2009})}\BibitemShut {NoStop}%
\bibitem [{\citenamefont {Coakley}\ and\ \citenamefont
  {Nyborg}(1978)}]{coakley1978cavitation}%
  \BibitemOpen
  \bibfield  {author} {\bibinfo {author} {\bibfnamefont {W.~T.}\ \bibnamefont
  {Coakley}}\ and\ \bibinfo {author} {\bibfnamefont {W.}~\bibnamefont
  {Nyborg}},\ }\href@noop {} {\bibfield  {journal} {\bibinfo  {journal}
  {Ultrasound: Its applications in medicine and biology}\ }\textbf {\bibinfo
  {volume} {3}},\ \bibinfo {pages} {77} (\bibinfo {year} {1978})}\BibitemShut
  {NoStop}%
\bibitem [{\citenamefont {Bruus}(2012)}]{bruus2012acoustofluidics}%
  \BibitemOpen
  \bibfield  {author} {\bibinfo {author} {\bibfnamefont {H.}~\bibnamefont
  {Bruus}},\ }\href@noop {} {\bibfield  {journal} {\bibinfo  {journal} {Lab on
  a Chip}\ }\textbf {\bibinfo {volume} {12}},\ \bibinfo {pages} {1014}
  (\bibinfo {year} {2012})}\BibitemShut {NoStop}%
\bibitem [{\citenamefont {Agarwal}\ \emph {et~al.}(2018)\citenamefont
  {Agarwal}, \citenamefont {Rallabandi},\ and\ \citenamefont
  {Hilgenfeldt}}]{agarwal2018inertial}%
  \BibitemOpen
  \bibfield  {author} {\bibinfo {author} {\bibfnamefont {S.}~\bibnamefont
  {Agarwal}}, \bibinfo {author} {\bibfnamefont {B.}~\bibnamefont
  {Rallabandi}},\ and\ \bibinfo {author} {\bibfnamefont {S.}~\bibnamefont
  {Hilgenfeldt}},\ }\href@noop {} {\bibfield  {journal} {\bibinfo  {journal}
  {Physical Review Fluids}\ }\textbf {\bibinfo {volume} {3}},\ \bibinfo {pages}
  {104201} (\bibinfo {year} {2018})}\BibitemShut {NoStop}%
\bibitem [{\citenamefont {Wang}\ \emph {et~al.}(2012)\citenamefont {Wang},
  \citenamefont {Jalikop},\ and\ \citenamefont
  {Hilgenfeldt}}]{wang2012efficient}%
  \BibitemOpen
  \bibfield  {author} {\bibinfo {author} {\bibfnamefont {C.}~\bibnamefont
  {Wang}}, \bibinfo {author} {\bibfnamefont {S.~V.}\ \bibnamefont {Jalikop}},\
  and\ \bibinfo {author} {\bibfnamefont {S.}~\bibnamefont {Hilgenfeldt}},\
  }\href@noop {} {\bibfield  {journal} {\bibinfo  {journal} {Biomicrofluidics}\
  }\textbf {\bibinfo {volume} {6}},\ \bibinfo {pages} {012801} (\bibinfo {year}
  {2012})}\BibitemShut {NoStop}%
\bibitem [{\citenamefont {Wang}\ \emph {et~al.}(2013)\citenamefont {Wang},
  \citenamefont {Rallabandi},\ and\ \citenamefont
  {Hilgenfeldt}}]{wang2013frequency}%
  \BibitemOpen
  \bibfield  {author} {\bibinfo {author} {\bibfnamefont {C.}~\bibnamefont
  {Wang}}, \bibinfo {author} {\bibfnamefont {B.}~\bibnamefont {Rallabandi}},\
  and\ \bibinfo {author} {\bibfnamefont {S.}~\bibnamefont {Hilgenfeldt}},\
  }\href@noop {} {\bibfield  {journal} {\bibinfo  {journal} {Physics of
  Fluids}\ }\textbf {\bibinfo {volume} {25}},\ \bibinfo {pages} {022002}
  (\bibinfo {year} {2013})}\BibitemShut {NoStop}%
\bibitem [{\citenamefont {Hashmi}\ \emph {et~al.}(2012)\citenamefont {Hashmi},
  \citenamefont {Yu}, \citenamefont {Reilly-Collette}, \citenamefont {Heiman},\
  and\ \citenamefont {Xu}}]{hashmi2012oscillating}%
  \BibitemOpen
  \bibfield  {author} {\bibinfo {author} {\bibfnamefont {A.}~\bibnamefont
  {Hashmi}}, \bibinfo {author} {\bibfnamefont {G.}~\bibnamefont {Yu}}, \bibinfo
  {author} {\bibfnamefont {M.}~\bibnamefont {Reilly-Collette}}, \bibinfo
  {author} {\bibfnamefont {G.}~\bibnamefont {Heiman}},\ and\ \bibinfo {author}
  {\bibfnamefont {J.}~\bibnamefont {Xu}},\ }\href@noop {} {\bibfield  {journal}
  {\bibinfo  {journal} {Lab on a Chip}\ }\textbf {\bibinfo {volume} {12}},\
  \bibinfo {pages} {4216} (\bibinfo {year} {2012})}\BibitemShut {NoStop}%
\bibitem [{\citenamefont {Chen}\ \emph {et~al.}(2016)\citenamefont {Chen},
  \citenamefont {Fang}, \citenamefont {Merritt}, \citenamefont {Strack},
  \citenamefont {Xu},\ and\ \citenamefont {Lee}}]{chen2016onset}%
  \BibitemOpen
  \bibfield  {author} {\bibinfo {author} {\bibfnamefont {Y.}~\bibnamefont
  {Chen}}, \bibinfo {author} {\bibfnamefont {Z.}~\bibnamefont {Fang}}, \bibinfo
  {author} {\bibfnamefont {B.}~\bibnamefont {Merritt}}, \bibinfo {author}
  {\bibfnamefont {D.}~\bibnamefont {Strack}}, \bibinfo {author} {\bibfnamefont
  {J.}~\bibnamefont {Xu}},\ and\ \bibinfo {author} {\bibfnamefont
  {S.}~\bibnamefont {Lee}},\ }\href@noop {} {\bibfield  {journal} {\bibinfo
  {journal} {Lab on a Chip}\ }\textbf {\bibinfo {volume} {16}},\ \bibinfo
  {pages} {3024} (\bibinfo {year} {2016})}\BibitemShut {NoStop}%
\bibitem [{\citenamefont {Longuet-Higgins}(1998)}]{lon98}%
  \BibitemOpen
  \bibfield  {author} {\bibinfo {author} {\bibfnamefont {M.~S.}\ \bibnamefont
  {Longuet-Higgins}},\ }\href@noop {} {\bibfield  {journal} {\bibinfo
  {journal} {Proceedings of the Royal Society of London. Series A:
  Mathematical, Physical and Engineering Sciences}\ }\textbf {\bibinfo {volume}
  {454}},\ \bibinfo {pages} {725} (\bibinfo {year} {1998})}\BibitemShut
  {NoStop}%
\bibitem [{\citenamefont {Rallabandi}(2020)}]{ral20_MRNote_preprint}%
  \BibitemOpen
  \bibfield  {author} {\bibinfo {author} {\bibfnamefont {B.}~\bibnamefont
  {Rallabandi}},\ }\href@noop {} {\  (\bibinfo {year} {2020})},\ \bibinfo
  {note} {preprint}\BibitemShut {NoStop}%
\bibitem [{\citenamefont {Saffman}(1965)}]{saffman1965lift}%
  \BibitemOpen
  \bibfield  {author} {\bibinfo {author} {\bibfnamefont {P.}~\bibnamefont
  {Saffman}},\ }\href@noop {} {\bibfield  {journal} {\bibinfo  {journal}
  {Journal of fluid mechanics}\ }\textbf {\bibinfo {volume} {22}},\ \bibinfo
  {pages} {385} (\bibinfo {year} {1965})}\BibitemShut {NoStop}%
\bibitem [{\citenamefont {Parthasarathy}\ \emph {et~al.}(2019)\citenamefont
  {Parthasarathy}, \citenamefont {Chan},\ and\ \citenamefont
  {Gazzola}}]{parthasarathy2019streaming}%
  \BibitemOpen
  \bibfield  {author} {\bibinfo {author} {\bibfnamefont {T.}~\bibnamefont
  {Parthasarathy}}, \bibinfo {author} {\bibfnamefont {F.~K.}\ \bibnamefont
  {Chan}},\ and\ \bibinfo {author} {\bibfnamefont {M.}~\bibnamefont
  {Gazzola}},\ }\href@noop {} {\bibfield  {journal} {\bibinfo  {journal}
  {Journal of Fluid Mechanics}\ }\textbf {\bibinfo {volume} {878}},\ \bibinfo
  {pages} {647} (\bibinfo {year} {2019})}\BibitemShut {NoStop}%
\bibitem [{\citenamefont {Bhosale}\ \emph {et~al.}(2020)\citenamefont
  {Bhosale}, \citenamefont {Parthasarathy},\ and\ \citenamefont
  {Gazzola}}]{bhosale_parthasarathy_gazzola_2020}%
  \BibitemOpen
  \bibfield  {author} {\bibinfo {author} {\bibfnamefont {Y.}~\bibnamefont
  {Bhosale}}, \bibinfo {author} {\bibfnamefont {T.}~\bibnamefont
  {Parthasarathy}},\ and\ \bibinfo {author} {\bibfnamefont {M.}~\bibnamefont
  {Gazzola}},\ }\href {https://doi.org/10.1017/jfm.2020.404} {\bibfield
  {journal} {\bibinfo  {journal} {Journal of Fluid Mechanics}\ }\textbf
  {\bibinfo {volume} {898}},\ \bibinfo {pages} {A13} (\bibinfo {year}
  {2020})}\BibitemShut {NoStop}%
\bibitem [{\citenamefont {Towns}\ \emph {et~al.}(2014)\citenamefont {Towns},
  \citenamefont {Cockerill}, \citenamefont {Dahan}, \citenamefont {Foster},
  \citenamefont {Gaither}, \citenamefont {Grimshaw}, \citenamefont {Hazlewood},
  \citenamefont {Lathrop}, \citenamefont {Lifka}, \citenamefont {Peterson},
  \citenamefont {Roskies}, \citenamefont {Scott},\ and\ \citenamefont
  {Wilkins-Diehr}}]{Towns:2014}%
  \BibitemOpen
  \bibfield  {author} {\bibinfo {author} {\bibfnamefont {J.}~\bibnamefont
  {Towns}}, \bibinfo {author} {\bibfnamefont {T.}~\bibnamefont {Cockerill}},
  \bibinfo {author} {\bibfnamefont {M.}~\bibnamefont {Dahan}}, \bibinfo
  {author} {\bibfnamefont {I.}~\bibnamefont {Foster}}, \bibinfo {author}
  {\bibfnamefont {K.}~\bibnamefont {Gaither}}, \bibinfo {author} {\bibfnamefont
  {A.}~\bibnamefont {Grimshaw}}, \bibinfo {author} {\bibfnamefont
  {V.}~\bibnamefont {Hazlewood}}, \bibinfo {author} {\bibfnamefont
  {S.}~\bibnamefont {Lathrop}}, \bibinfo {author} {\bibfnamefont
  {D.}~\bibnamefont {Lifka}}, \bibinfo {author} {\bibfnamefont {G.~D.}\
  \bibnamefont {Peterson}}, \bibinfo {author} {\bibfnamefont {R.}~\bibnamefont
  {Roskies}}, \bibinfo {author} {\bibfnamefont {J.~R.}\ \bibnamefont {Scott}},\
  and\ \bibinfo {author} {\bibfnamefont {N.}~\bibnamefont {Wilkins-Diehr}},\
  }\href {https://doi.org/10.1109/MCSE.2014.80} {\bibfield  {journal} {\bibinfo
   {journal} {Computing in Science \& Engineering}\ }\textbf {\bibinfo {volume}
  {16}},\ \bibinfo {pages} {62} (\bibinfo {year} {2014})}\BibitemShut {NoStop}%
\bibitem [{\citenamefont {Landau}\ and\ \citenamefont
  {Lifshitz}(1959)}]{landau1959course}%
  \BibitemOpen
  \bibfield  {author} {\bibinfo {author} {\bibfnamefont {L.~D.}\ \bibnamefont
  {Landau}}\ and\ \bibinfo {author} {\bibfnamefont {E.}~\bibnamefont
  {Lifshitz}},\ }\href@noop {} {\emph {\bibinfo {title} {Course of Theoretical
  Physics Vol. 6 Fluid Mechanies}}}\ (\bibinfo  {publisher} {Pergamon Press},\
  \bibinfo {year} {1959})\BibitemShut {NoStop}%
\bibitem [{\citenamefont {Pozrikidis}\ \emph {et~al.}(1992)\citenamefont
  {Pozrikidis} \emph {et~al.}}]{pozrikidis1992boundary}%
  \BibitemOpen
  \bibfield  {author} {\bibinfo {author} {\bibfnamefont {C.}~\bibnamefont
  {Pozrikidis}} \emph {et~al.},\ }\href@noop {} {\emph {\bibinfo {title}
  {Boundary integral and singularity methods for linearized viscous flow}}}\
  (\bibinfo  {publisher} {Cambridge University Press},\ \bibinfo {year}
  {1992})\BibitemShut {NoStop}%
\bibitem [{\citenamefont {Danilov}\ and\ \citenamefont
  {Mironov}(2000)}]{danilov2000mean}%
  \BibitemOpen
  \bibfield  {author} {\bibinfo {author} {\bibfnamefont {S.}~\bibnamefont
  {Danilov}}\ and\ \bibinfo {author} {\bibfnamefont {M.}~\bibnamefont
  {Mironov}},\ }\href@noop {} {\bibfield  {journal} {\bibinfo  {journal} {The
  Journal of the Acoustical Society of America}\ }\textbf {\bibinfo {volume}
  {107}},\ \bibinfo {pages} {143} (\bibinfo {year} {2000})}\BibitemShut
  {NoStop}%
\bibitem [{\citenamefont {Nadim}\ and\ \citenamefont
  {Stone}(1991)}]{nadim1991motion}%
  \BibitemOpen
  \bibfield  {author} {\bibinfo {author} {\bibfnamefont {A.}~\bibnamefont
  {Nadim}}\ and\ \bibinfo {author} {\bibfnamefont {H.~A.}\ \bibnamefont
  {Stone}},\ }\href@noop {} {\bibfield  {journal} {\bibinfo  {journal} {Studies
  in Applied Mathematics}\ }\textbf {\bibinfo {volume} {85}},\ \bibinfo {pages}
  {53} (\bibinfo {year} {1991})}\BibitemShut {NoStop}%
\bibitem [{\citenamefont {Basset}(1888)}]{basset1888treatise}%
  \BibitemOpen
  \bibfield  {author} {\bibinfo {author} {\bibfnamefont {A.~B.}\ \bibnamefont
  {Basset}},\ }\href@noop {} {\emph {\bibinfo {title} {A treatise on
  hydrodynamics: with numerous examples}}},\ Vol.~\bibinfo {volume} {2}\
  (\bibinfo  {publisher} {Deighton, Bell and Company},\ \bibinfo {year}
  {1888})\BibitemShut {NoStop}%
\bibitem [{\citenamefont {Gazzola}\ \emph {et~al.}(2011)\citenamefont
  {Gazzola}, \citenamefont {Chatelain}, \citenamefont {Van~Rees},\ and\
  \citenamefont {Koumoutsakos}}]{gazzola2011simulations}%
  \BibitemOpen
  \bibfield  {author} {\bibinfo {author} {\bibfnamefont {M.}~\bibnamefont
  {Gazzola}}, \bibinfo {author} {\bibfnamefont {P.}~\bibnamefont {Chatelain}},
  \bibinfo {author} {\bibfnamefont {W.~M.}\ \bibnamefont {Van~Rees}},\ and\
  \bibinfo {author} {\bibfnamefont {P.}~\bibnamefont {Koumoutsakos}},\
  }\href@noop {} {\bibfield  {journal} {\bibinfo  {journal} {Journal of
  Computational Physics}\ }\textbf {\bibinfo {volume} {230}},\ \bibinfo {pages}
  {7093} (\bibinfo {year} {2011})}\BibitemShut {NoStop}%
\bibitem [{\citenamefont {Gazzola}\ \emph
  {et~al.}(2012{\natexlab{a}})\citenamefont {Gazzola}, \citenamefont {Mimeau},
  \citenamefont {Tchieu},\ and\ \citenamefont
  {Koumoutsakos}}]{gazzola2012flow}%
  \BibitemOpen
  \bibfield  {author} {\bibinfo {author} {\bibfnamefont {M.}~\bibnamefont
  {Gazzola}}, \bibinfo {author} {\bibfnamefont {C.}~\bibnamefont {Mimeau}},
  \bibinfo {author} {\bibfnamefont {A.~A.}\ \bibnamefont {Tchieu}},\ and\
  \bibinfo {author} {\bibfnamefont {P.}~\bibnamefont {Koumoutsakos}},\
  }\href@noop {} {\bibfield  {journal} {\bibinfo  {journal} {Physics of
  fluids}\ }\textbf {\bibinfo {volume} {24}},\ \bibinfo {pages} {043103}
  (\bibinfo {year} {2012}{\natexlab{a}})}\BibitemShut {NoStop}%
\bibitem [{\citenamefont {Gazzola}\ \emph
  {et~al.}(2012{\natexlab{b}})\citenamefont {Gazzola}, \citenamefont
  {Van~Rees},\ and\ \citenamefont {Koumoutsakos}}]{gazzola2012c}%
  \BibitemOpen
  \bibfield  {author} {\bibinfo {author} {\bibfnamefont {M.}~\bibnamefont
  {Gazzola}}, \bibinfo {author} {\bibfnamefont {W.~M.}\ \bibnamefont
  {Van~Rees}},\ and\ \bibinfo {author} {\bibfnamefont {P.}~\bibnamefont
  {Koumoutsakos}},\ }\href@noop {} {\bibfield  {journal} {\bibinfo  {journal}
  {Journal of Fluid Mechanics}\ }\textbf {\bibinfo {volume} {698}},\ \bibinfo
  {pages} {5} (\bibinfo {year} {2012}{\natexlab{b}})}\BibitemShut {NoStop}%
\bibitem [{\citenamefont {Gazzola}\ \emph {et~al.}(2014)\citenamefont
  {Gazzola}, \citenamefont {Hejazialhosseini},\ and\ \citenamefont
  {Koumoutsakos}}]{gazzola2014reinforcement}%
  \BibitemOpen
  \bibfield  {author} {\bibinfo {author} {\bibfnamefont {M.}~\bibnamefont
  {Gazzola}}, \bibinfo {author} {\bibfnamefont {B.}~\bibnamefont
  {Hejazialhosseini}},\ and\ \bibinfo {author} {\bibfnamefont {P.}~\bibnamefont
  {Koumoutsakos}},\ }\href@noop {} {\bibfield  {journal} {\bibinfo  {journal}
  {SIAM Journal on Scientific Computing}\ }\textbf {\bibinfo {volume} {36}},\
  \bibinfo {pages} {B622} (\bibinfo {year} {2014})}\BibitemShut {NoStop}%
\bibitem [{\citenamefont {Gazzola}\ \emph {et~al.}(2016)\citenamefont
  {Gazzola}, \citenamefont {Tchieu}, \citenamefont {Alexeev}, \citenamefont
  {de~Brauer},\ and\ \citenamefont {Koumoutsakos}}]{gazzola2016learning}%
  \BibitemOpen
  \bibfield  {author} {\bibinfo {author} {\bibfnamefont {M.}~\bibnamefont
  {Gazzola}}, \bibinfo {author} {\bibfnamefont {A.~A.}\ \bibnamefont {Tchieu}},
  \bibinfo {author} {\bibfnamefont {D.}~\bibnamefont {Alexeev}}, \bibinfo
  {author} {\bibfnamefont {A.}~\bibnamefont {de~Brauer}},\ and\ \bibinfo
  {author} {\bibfnamefont {P.}~\bibnamefont {Koumoutsakos}},\ }\href@noop {}
  {\bibfield  {journal} {\bibinfo  {journal} {Journal of Fluid Mechanics}\
  }\textbf {\bibinfo {volume} {789}},\ \bibinfo {pages} {726} (\bibinfo {year}
  {2016})}\BibitemShut {NoStop}%
\end{thebibliography}%


\providecommand{\noopsort}[1]{}\providecommand{\singleletter}[1]{#1}%
\begin{thebibliography}{18}%
\makeatletter
\providecommand \@ifxundefined [1]{%
 \@ifx{#1\undefined}
}%
\providecommand \@ifnum [1]{%
 \ifnum #1\expandafter \@firstoftwo
 \else \expandafter \@secondoftwo
 \fi
}%
\providecommand \@ifx [1]{%
 \ifx #1\expandafter \@firstoftwo
 \else \expandafter \@secondoftwo
 \fi
}%
\providecommand \natexlab [1]{#1}%
\providecommand \enquote  [1]{``#1''}%
\providecommand \bibnamefont  [1]{#1}%
\providecommand \bibfnamefont [1]{#1}%
\providecommand \citenamefont [1]{#1}%
\providecommand \href@noop [0]{\@secondoftwo}%
\providecommand \href [0]{\begingroup \@sanitize@url \@href}%
\providecommand \@href[1]{\@@startlink{#1}\@@href}%
\providecommand \@@href[1]{\endgroup#1\@@endlink}%
\providecommand \@sanitize@url [0]{\catcode `\\12\catcode `\$12\catcode
  `\&12\catcode `\#12\catcode `\^12\catcode `\_12\catcode `\%12\relax}%
\providecommand \@@startlink[1]{}%
\providecommand \@@endlink[0]{}%
\providecommand \url  [0]{\begingroup\@sanitize@url \@url }%
\providecommand \@url [1]{\endgroup\@href {#1}{\urlprefix }}%
\providecommand \urlprefix  [0]{URL }%
\providecommand \Eprint [0]{\href }%
\providecommand \doibase [0]{https://doi.org/}%
\providecommand \selectlanguage [0]{\@gobble}%
\providecommand \bibinfo  [0]{\@secondoftwo}%
\providecommand \bibfield  [0]{\@secondoftwo}%
\providecommand \translation [1]{[#1]}%
\providecommand \BibitemOpen [0]{}%
\providecommand \bibitemStop [0]{}%
\providecommand \bibitemNoStop [0]{.\EOS\space}%
\providecommand \EOS [0]{\spacefactor3000\relax}%
\providecommand \BibitemShut  [1]{\csname bibitem#1\endcsname}%
\let\auto@bib@innerbib\@empty
\bibitem [{\citenamefont {Maxey}\ and\ \citenamefont
  {Riley}(1983)}]{maxey1983equation}%
  \BibitemOpen
  \bibfield  {author} {\bibinfo {author} {\bibfnamefont {M.~R.}\ \bibnamefont
  {Maxey}}\ and\ \bibinfo {author} {\bibfnamefont {J.~J.}\ \bibnamefont
  {Riley}},\ }\bibfield  {title} {\bibinfo {title} {Equation of motion for a
  small rigid sphere in a nonuniform flow},\ }\href
  {https://doi.org/10.1063/1.864230} {\bibfield  {journal} {\bibinfo  {journal}
  {The Physics of Fluids}\ }\textbf {\bibinfo {volume} {26}},\ \bibinfo {pages}
  {883} (\bibinfo {year} {1983})}\BibitemShut {NoStop}%
\bibitem [{\citenamefont {Cox}\ and\ \citenamefont
  {Brenner}(1968)}]{cox1968lateral}%
  \BibitemOpen
  \bibfield  {author} {\bibinfo {author} {\bibfnamefont {R.}~\bibnamefont
  {Cox}}\ and\ \bibinfo {author} {\bibfnamefont {H.}~\bibnamefont {Brenner}},\
  }\bibfield  {title} {\bibinfo {title} {The lateral migration of solid
  particles in poiseuille flow---i theory},\ }\href@noop {} {\bibfield
  {journal} {\bibinfo  {journal} {Chemical Engineering Science}\ }\textbf
  {\bibinfo {volume} {23}},\ \bibinfo {pages} {147} (\bibinfo {year}
  {1968})}\BibitemShut {NoStop}%
\bibitem [{\citenamefont {Ho}\ and\ \citenamefont
  {Leal}(1974)}]{ho1974inertial}%
  \BibitemOpen
  \bibfield  {author} {\bibinfo {author} {\bibfnamefont {B.}~\bibnamefont
  {Ho}}\ and\ \bibinfo {author} {\bibfnamefont {L.}~\bibnamefont {Leal}},\
  }\bibfield  {title} {\bibinfo {title} {Inertial migration of rigid spheres in
  two-dimensional unidirectional flows},\ }\href@noop {} {\bibfield  {journal}
  {\bibinfo  {journal} {Journal of fluid mechanics}\ }\textbf {\bibinfo
  {volume} {65}},\ \bibinfo {pages} {365} (\bibinfo {year} {1974})}\BibitemShut
  {NoStop}%
\bibitem [{\citenamefont {Hood}\ \emph {et~al.}(2015)\citenamefont {Hood},
  \citenamefont {Lee},\ and\ \citenamefont {Roper}}]{hood2015inertial}%
  \BibitemOpen
  \bibfield  {author} {\bibinfo {author} {\bibfnamefont {K.}~\bibnamefont
  {Hood}}, \bibinfo {author} {\bibfnamefont {S.}~\bibnamefont {Lee}},\ and\
  \bibinfo {author} {\bibfnamefont {M.}~\bibnamefont {Roper}},\ }\bibfield
  {title} {\bibinfo {title} {Inertial migration of a rigid sphere in
  three-dimensional poiseuille flow},\ }\href@noop {} {\bibfield  {journal}
  {\bibinfo  {journal} {Journal of Fluid Mechanics}\ }\textbf {\bibinfo
  {volume} {765}},\ \bibinfo {pages} {452} (\bibinfo {year}
  {2015})}\BibitemShut {NoStop}%
\bibitem [{\citenamefont {Landau}\ and\ \citenamefont
  {Lifshitz}(1959)}]{landau1959course}%
  \BibitemOpen
  \bibfield  {author} {\bibinfo {author} {\bibfnamefont {L.~D.}\ \bibnamefont
  {Landau}}\ and\ \bibinfo {author} {\bibfnamefont {E.}~\bibnamefont
  {Lifshitz}},\ }\href@noop {} {\emph {\bibinfo {title} {Course of Theoretical
  Physics Vol. 6 Fluid Mechanies}}}\ (\bibinfo  {publisher} {Pergamon Press},\
  \bibinfo {year} {1959})\BibitemShut {NoStop}%
\bibitem [{\citenamefont {Pozrikidis}\ \emph {et~al.}(1992)\citenamefont
  {Pozrikidis} \emph {et~al.}}]{pozrikidis1992boundary}%
  \BibitemOpen
  \bibfield  {author} {\bibinfo {author} {\bibfnamefont {C.}~\bibnamefont
  {Pozrikidis}} \emph {et~al.},\ }\href@noop {} {\emph {\bibinfo {title}
  {Boundary integral and singularity methods for linearized viscous flow}}}\
  (\bibinfo  {publisher} {Cambridge University Press},\ \bibinfo {year}
  {1992})\BibitemShut {NoStop}%
\bibitem [{\citenamefont {Lovalenti}\ and\ \citenamefont
  {Brady}(1993)}]{lovalenti1993hydrodynamic}%
  \BibitemOpen
  \bibfield  {author} {\bibinfo {author} {\bibfnamefont {P.~M.}\ \bibnamefont
  {Lovalenti}}\ and\ \bibinfo {author} {\bibfnamefont {J.~F.}\ \bibnamefont
  {Brady}},\ }\bibfield  {title} {\bibinfo {title} {The hydrodynamic force on a
  rigid particle undergoing arbitrary time-dependent motion at small reynolds
  number},\ }\href@noop {} {\bibfield  {journal} {\bibinfo  {journal} {Journal
  of Fluid Mechanics}\ }\textbf {\bibinfo {volume} {256}},\ \bibinfo {pages}
  {561} (\bibinfo {year} {1993})}\BibitemShut {NoStop}%
\bibitem [{\citenamefont {Stone}\ \emph {et~al.}(2001)\citenamefont {Stone},
  \citenamefont {Brady},\ and\ \citenamefont {Lovalenti}}]{stone2001inertial}%
  \BibitemOpen
  \bibfield  {author} {\bibinfo {author} {\bibfnamefont {H.}~\bibnamefont
  {Stone}}, \bibinfo {author} {\bibfnamefont {J.}~\bibnamefont {Brady}},\ and\
  \bibinfo {author} {\bibfnamefont {P.}~\bibnamefont {Lovalenti}},\ }\bibfield
  {title} {\bibinfo {title} {Inertial effects on the rheology of suspensions
  and on the motion of individual particles},\ }\href@noop {} {\bibfield
  {journal} {\bibinfo  {journal} {preprint}\ } (\bibinfo {year}
  {2001})}\BibitemShut {NoStop}%
\bibitem [{\citenamefont {Rallabandi}(2020)}]{ral20_MRNote_preprint}%
  \BibitemOpen
  \bibfield  {author} {\bibinfo {author} {\bibfnamefont {B.}~\bibnamefont
  {Rallabandi}},\ }\bibfield  {title} {\bibinfo {title} {A note on the
  {M}axey--{R}iley equation in nonuniform flows},\ }\href@noop {} {\  (\bibinfo
  {year} {2020})},\ \bibinfo {note} {preprint}\BibitemShut {NoStop}%
\bibitem [{\citenamefont {Danilov}\ and\ \citenamefont
  {Mironov}(2000)}]{danilov2000mean}%
  \BibitemOpen
  \bibfield  {author} {\bibinfo {author} {\bibfnamefont {S.}~\bibnamefont
  {Danilov}}\ and\ \bibinfo {author} {\bibfnamefont {M.}~\bibnamefont
  {Mironov}},\ }\bibfield  {title} {\bibinfo {title} {Mean force on a small
  sphere in a sound field in a viscous fluid},\ }\href@noop {} {\bibfield
  {journal} {\bibinfo  {journal} {The Journal of the Acoustical Society of
  America}\ }\textbf {\bibinfo {volume} {107}},\ \bibinfo {pages} {143}
  (\bibinfo {year} {2000})}\BibitemShut {NoStop}%
\bibitem [{\citenamefont {Gazzola}\ \emph {et~al.}(2011)\citenamefont
  {Gazzola}, \citenamefont {Chatelain}, \citenamefont {Van~Rees},\ and\
  \citenamefont {Koumoutsakos}}]{gazzola2011simulations}%
  \BibitemOpen
  \bibfield  {author} {\bibinfo {author} {\bibfnamefont {M.}~\bibnamefont
  {Gazzola}}, \bibinfo {author} {\bibfnamefont {P.}~\bibnamefont {Chatelain}},
  \bibinfo {author} {\bibfnamefont {W.~M.}\ \bibnamefont {Van~Rees}},\ and\
  \bibinfo {author} {\bibfnamefont {P.}~\bibnamefont {Koumoutsakos}},\
  }\bibfield  {title} {\bibinfo {title} {Simulations of single and multiple
  swimmers with non-divergence free deforming geometries},\ }\href@noop {}
  {\bibfield  {journal} {\bibinfo  {journal} {Journal of Computational
  Physics}\ }\textbf {\bibinfo {volume} {230}},\ \bibinfo {pages} {7093}
  (\bibinfo {year} {2011})}\BibitemShut {NoStop}%
\bibitem [{\citenamefont {Gazzola}\ \emph
  {et~al.}(2012{\natexlab{a}})\citenamefont {Gazzola}, \citenamefont {Mimeau},
  \citenamefont {Tchieu},\ and\ \citenamefont
  {Koumoutsakos}}]{gazzola2012flow}%
  \BibitemOpen
  \bibfield  {author} {\bibinfo {author} {\bibfnamefont {M.}~\bibnamefont
  {Gazzola}}, \bibinfo {author} {\bibfnamefont {C.}~\bibnamefont {Mimeau}},
  \bibinfo {author} {\bibfnamefont {A.~A.}\ \bibnamefont {Tchieu}},\ and\
  \bibinfo {author} {\bibfnamefont {P.}~\bibnamefont {Koumoutsakos}},\
  }\bibfield  {title} {\bibinfo {title} {Flow mediated interactions between two
  cylinders at finite re numbers},\ }\href@noop {} {\bibfield  {journal}
  {\bibinfo  {journal} {Physics of fluids}\ }\textbf {\bibinfo {volume} {24}},\
  \bibinfo {pages} {043103} (\bibinfo {year} {2012}{\natexlab{a}})}\BibitemShut
  {NoStop}%
\bibitem [{\citenamefont {Gazzola}\ \emph
  {et~al.}(2012{\natexlab{b}})\citenamefont {Gazzola}, \citenamefont
  {Van~Rees},\ and\ \citenamefont {Koumoutsakos}}]{gazzola2012c}%
  \BibitemOpen
  \bibfield  {author} {\bibinfo {author} {\bibfnamefont {M.}~\bibnamefont
  {Gazzola}}, \bibinfo {author} {\bibfnamefont {W.~M.}\ \bibnamefont
  {Van~Rees}},\ and\ \bibinfo {author} {\bibfnamefont {P.}~\bibnamefont
  {Koumoutsakos}},\ }\bibfield  {title} {\bibinfo {title} {C-start: optimal
  start of larval fish},\ }\href@noop {} {\bibfield  {journal} {\bibinfo
  {journal} {Journal of Fluid Mechanics}\ }\textbf {\bibinfo {volume} {698}},\
  \bibinfo {pages} {5} (\bibinfo {year} {2012}{\natexlab{b}})}\BibitemShut
  {NoStop}%
\bibitem [{\citenamefont {Gazzola}\ \emph {et~al.}(2014)\citenamefont
  {Gazzola}, \citenamefont {Hejazialhosseini},\ and\ \citenamefont
  {Koumoutsakos}}]{gazzola2014reinforcement}%
  \BibitemOpen
  \bibfield  {author} {\bibinfo {author} {\bibfnamefont {M.}~\bibnamefont
  {Gazzola}}, \bibinfo {author} {\bibfnamefont {B.}~\bibnamefont
  {Hejazialhosseini}},\ and\ \bibinfo {author} {\bibfnamefont {P.}~\bibnamefont
  {Koumoutsakos}},\ }\bibfield  {title} {\bibinfo {title} {Reinforcement
  learning and wavelet adapted vortex methods for simulations of self-propelled
  swimmers},\ }\href@noop {} {\bibfield  {journal} {\bibinfo  {journal} {SIAM
  Journal on Scientific Computing}\ }\textbf {\bibinfo {volume} {36}},\
  \bibinfo {pages} {B622} (\bibinfo {year} {2014})}\BibitemShut {NoStop}%
\bibitem [{\citenamefont {Gazzola}\ \emph {et~al.}(2016)\citenamefont
  {Gazzola}, \citenamefont {Tchieu}, \citenamefont {Alexeev}, \citenamefont
  {de~Brauer},\ and\ \citenamefont {Koumoutsakos}}]{gazzola2016learning}%
  \BibitemOpen
  \bibfield  {author} {\bibinfo {author} {\bibfnamefont {M.}~\bibnamefont
  {Gazzola}}, \bibinfo {author} {\bibfnamefont {A.~A.}\ \bibnamefont {Tchieu}},
  \bibinfo {author} {\bibfnamefont {D.}~\bibnamefont {Alexeev}}, \bibinfo
  {author} {\bibfnamefont {A.}~\bibnamefont {de~Brauer}},\ and\ \bibinfo
  {author} {\bibfnamefont {P.}~\bibnamefont {Koumoutsakos}},\ }\bibfield
  {title} {\bibinfo {title} {Learning to school in the presence of hydrodynamic
  interactions},\ }\href@noop {} {\bibfield  {journal} {\bibinfo  {journal}
  {Journal of Fluid Mechanics}\ }\textbf {\bibinfo {volume} {789}},\ \bibinfo
  {pages} {726} (\bibinfo {year} {2016})}\BibitemShut {NoStop}%
\bibitem [{\citenamefont {Parthasarathy}\ \emph {et~al.}(2019)\citenamefont
  {Parthasarathy}, \citenamefont {Chan},\ and\ \citenamefont
  {Gazzola}}]{parthasarathy2019streaming}%
  \BibitemOpen
  \bibfield  {author} {\bibinfo {author} {\bibfnamefont {T.}~\bibnamefont
  {Parthasarathy}}, \bibinfo {author} {\bibfnamefont {F.~K.}\ \bibnamefont
  {Chan}},\ and\ \bibinfo {author} {\bibfnamefont {M.}~\bibnamefont
  {Gazzola}},\ }\bibfield  {title} {\bibinfo {title} {Streaming-enhanced
  flow-mediated transport},\ }\href@noop {} {\bibfield  {journal} {\bibinfo
  {journal} {Journal of Fluid Mechanics}\ }\textbf {\bibinfo {volume} {878}},\
  \bibinfo {pages} {647} (\bibinfo {year} {2019})}\BibitemShut {NoStop}%
\bibitem [{\citenamefont {Bhosale}\ \emph {et~al.}(2020)\citenamefont
  {Bhosale}, \citenamefont {Parthasarathy},\ and\ \citenamefont
  {Gazzola}}]{bhosale_parthasarathy_gazzola_2020}%
  \BibitemOpen
  \bibfield  {author} {\bibinfo {author} {\bibfnamefont {Y.}~\bibnamefont
  {Bhosale}}, \bibinfo {author} {\bibfnamefont {T.}~\bibnamefont
  {Parthasarathy}},\ and\ \bibinfo {author} {\bibfnamefont {M.}~\bibnamefont
  {Gazzola}},\ }\bibfield  {title} {\bibinfo {title} {Shape curvature effects
  in viscous streaming},\ }\href {https://doi.org/10.1017/jfm.2020.404}
  {\bibfield  {journal} {\bibinfo  {journal} {Journal of Fluid Mechanics}\
  }\textbf {\bibinfo {volume} {898}},\ \bibinfo {pages} {A13} (\bibinfo {year}
  {2020})}\BibitemShut {NoStop}%
\bibitem [{\citenamefont {Sbalzarini}\ \emph {et~al.}(2006)\citenamefont
  {Sbalzarini}, \citenamefont {Walther}, \citenamefont {Bergdorf},
  \citenamefont {Hieber}, \citenamefont {Kotsalis},\ and\ \citenamefont
  {Koumoutsakos}}]{sbalzarini2006ppm}%
  \BibitemOpen
  \bibfield  {author} {\bibinfo {author} {\bibfnamefont {I.~F.}\ \bibnamefont
  {Sbalzarini}}, \bibinfo {author} {\bibfnamefont {J.~H.}\ \bibnamefont
  {Walther}}, \bibinfo {author} {\bibfnamefont {M.}~\bibnamefont {Bergdorf}},
  \bibinfo {author} {\bibfnamefont {S.~E.}\ \bibnamefont {Hieber}}, \bibinfo
  {author} {\bibfnamefont {E.~M.}\ \bibnamefont {Kotsalis}},\ and\ \bibinfo
  {author} {\bibfnamefont {P.}~\bibnamefont {Koumoutsakos}},\ }\bibfield
  {title} {\bibinfo {title} {Ppm--a highly efficient parallel particle--mesh
  library for the simulation of continuum systems},\ }\href@noop {} {\bibfield
  {journal} {\bibinfo  {journal} {Journal of Computational Physics}\ }\textbf
  {\bibinfo {volume} {215}},\ \bibinfo {pages} {566} (\bibinfo {year}
  {2006})}\BibitemShut {NoStop}%
\end{thebibliography}%
\end{document}


\sloppy
	\allowdisplaybreaks
	
	\title{Supplementary Information: An unrecognized force in inertial microfluidics}
	\author{Siddhansh Agarwal}
	\author{Fan Kiat Chan}
	\affiliation{Mechanical Science and Engineering, University of Illinois, Urbana-Champaign, Illinois 61801, USA}
	\author{Bhargav Rallabandi}
	\affiliation{Mechanical Engineering, University of California, Riverside, USA}
	\author{Mattia Gazzola}
	\affiliation{Mechanical Science and Engineering, University of Illinois, Urbana-Champaign, Illinois 61801, USA}
	\affiliation{National Center for Supercomputing Applications, University of Illinois, Urbana-Champaign, Illinois 61801, USA}
	\affiliation{Carl R.\ Woese Institute for Genomic Biology, University of Illinois, Urbana-Champaign, Illinois 61801, USA}
	\author{Sascha Hilgenfeldt}%
	\email{sascha@illinois.edu}
	\affiliation{Mechanical Science and Engineering, University of Illinois, Urbana-Champaign, Illinois 61801, USA}
	\maketitle
	\section{Theoretical Formalism}
	In order to systematically account for the inertial forces on a sphere of radius $a_p$ centered at $\bm{r}_p$ moving with with velocity $\bm{u}_p$ (neglecting effects of rotation) and exposed to a known (lab-frame) background undisturbed flow $\bar{\bm{u}}$, we split the Navier--Stokes equations that govern the flow field into an undisturbed flow $\boldsymbol{w}^{(0)}=\bar{\bm{u}}-\bm{u}_p$ and a disturbance flow $\boldsymbol{w}^{(1)}$ (we adopt the same notation as \cite{maxey1983equation}). Then, in a particle-centered (moving) coordinate system, we have
	\begin{subequations}
		\begin{align}
			\nabla^{2} \boldsymbol{w}^{(0)}-\nabla p^{(0)} =&3\lambda \frac{\partial \boldsymbol{w}^{(0)}}{\partial t}+\operatorname{Re}_{p}\left(\boldsymbol{w}^{(0)} \cdot \nabla\boldsymbol{w}^{(0)}\right),\label{undistflow} \\
			\nabla^{2} \boldsymbol{w}^{(1)}-\nabla p^{(1)} =&3\lambda \frac{\partial \boldsymbol{w}^{(1)}}{\partial t}+\operatorname{Re}_{p}\bigg[\left(\bm{\bar{u}}-\bm{u}_p\right) \cdot \nabla \boldsymbol{w}^{(1)}+\boldsymbol{w}^{(1)}\cdot \nabla \bm{\bar{u}} +\boldsymbol{w}^{(1)}\cdot \nabla\boldsymbol{w}^{(1)}\bigg], \label{distflow}\\
			\boldsymbol{\nabla} \cdot \boldsymbol{w}^{(0)} =&0, \quad \boldsymbol{\nabla} \cdot \boldsymbol{w}^{(1)} =0, \\
			\boldsymbol{w}^{(1)} =& \bm{u}_p-\bm{\bar{u}} \quad \text { on } r=1 \quad \text{and} \quad \boldsymbol{w}^{(1)} = 0 \quad \text { as } r \rightarrow \infty,
		\end{align}
	\end{subequations}
	where $\operatorname{Re}_p= U^*a_p/\nu$ is the particle Reynolds number. Quantities in these equations are non-dimensionalized by scaling velocities with $U^*$, lengths with $a_p$, pressure with $\mu U^*/a_p$, and time by $\omega^{-1}$.
	
	The force contribution from the undisturbed flow is $\bm{F}^{(0)}=(F_S/6\pi)\oint_S \bm{n}\cdot \bsigma^{(0)} dS$, like in the original Maxey--Riley (MR) formalism \cite{maxey1983equation}, where $ \bsigma^{(0)}=-p^{(0)}\bm{I} + \nabla \boldsymbol{w}^{(0)} + \left(\nabla \boldsymbol{w}^{(0)}\right)^T$ is the stress tensor associated with the undisturbed flow field $\boldsymbol{w}^{(0)}$, and $F_S/6\pi=\nu \rho a_p U^*$ is the Stokes drag scale.
	The force contribution at the disturbance flow order is given by $\bm{F}^{(1)}=(F_S/6\pi)\oint_S \bm{n}\cdot \bsigma^{(1)} dS$, where $ \bsigma^{(1)}=-p^{(1)}\bm{I} + \nabla \boldsymbol{w}^{(1)} + \left(\nabla \boldsymbol{w}^{(1)}\right)^T$ is the stress tensor associated with the disturbance flow field $\boldsymbol{w}^{(1)}$. The corresponding (dimensional) equation of motion for the particle then reads
	\begin{align}
		m_p \frac{d \bm{U}_p}{dt} &= \bm{F}^{(0)}+\bm{F}^{(1)}.
	\end{align}
	Note that everything up to this point is exact and no assumptions have been made. MR \cite{maxey1983equation} make the unsteady Stokes flow approximation in \eqref{distflow} by setting $\operatorname{Re}_p = 0$, and compute $\bm{F}^{(1)}$ without explicitly evaluating the disturbance flow, using a symmetry relation. While this assumption is plausible in many traditional microfluidic flow situations, fast oscillatory particle motion can give rise to large disturbance flow gradients so that the inertial terms on the RHS of \eqref{distflow} are not necessarily negligible compared to the viscous diffusion term (typically $\operatorname{Re}_{p}\sim \mathcal{O}(1)$, as in the experiment described in Fig.~1 of the main text).
	\subsection{Small $Re_p$ expansion}
	In order to make analytical progress, following \cite{cox1968lateral,ho1974inertial,hood2015inertial}, we expand $\boldsymbol{w}^{(1)}$, $p^{(1)}$, $\bm{r}_p$, $\bm{u}_p$ and $\bsigma^{(1)}$ (and consequently $\bm{F}^{(1)}$) in a regular asymptotic expansion for small $\operatorname{Re}_p$,
	\begin{subequations}
		\begin{align}
			\boldsymbol{w}^{(1)} &= \boldsymbol{w}^{(1)}_0 + \operatorname{Re}_p \boldsymbol{w}^{(1)}_1 + \dots,\\
			p^{(1)} &= p^{(1)}_0 +\operatorname{Re}_p p^{(1)}_1 +\dots,\\
			\bm{r}_p &= \bm{r}_{p_0} + \operatorname{Re}_p \bm{r}_{p_1}  + \dots ,\\
			\bm{u}_p &= \bm{u}_{p_0} + \operatorname{Re}_p \bm{u}_{p_1}  + \dots ,\\
			\bsigma^{(1)}&=\bsigma^{(1)}_0+\operatorname{Re}_p \bsigma^{(1)}_1 + \dots,\\
			\bm{F}^{(1)}&=\bm{F}^{(1)}_0+\operatorname{Re}_p \bm{F}^{(1)}_1 + \dots .
		\end{align}
	\end{subequations}
	The leading-order equations for ($\boldsymbol{w}^{(1)}_0$, $p^{(1)}_0$) are unsteady Stokes,
	\begin{subequations}
		\begin{align}
			\nabla^{2} \boldsymbol{w}^{(1)}_0-\nabla p^{(1)}_0 &=3\lambda \frac{\partial \boldsymbol{w}^{(1)}_0}{\partial t}, \\
			\boldsymbol{\nabla} \cdot \boldsymbol{w}^{(1)}_0 &=0, \\
			\boldsymbol{w}^{(1)}_0 &= \bm{u}_{p_0}-\bm{\bar{u}} \quad \text { on } \bm{r}=1, \label{eqn:w10bc}\\
			\boldsymbol{w}^{(1)}_0 &= 0 \quad \text { as } \bm{r} \rightarrow \infty.
		\end{align}\label{eqn:w01}\noindent
	\end{subequations}
	We note that in the original derivation of MR \cite{maxey1983equation}, a symmetry relation was used at this order to compute $\bm{F}^{(1)}_0$ without explicitly solving for $\boldsymbol{w}^{(1)}_0$. However, since we are interested in computing the force contribution at $\mathcal{O}(\operatorname{Re}_p)$, we need an explicit solution for the leading-order disturbance flow $\boldsymbol{w}^{(1)}_0$. To obtain explicit results, as stated in the main text, we expand the background flow field $\bm{\bar{u}}$ around the leading-order particle position $\bm{r}_{p_0}$ into spatial moments of alternating symmetry,
	\begin{align}
		\bm{\bar{u}}=\bm{\bar{u}}|_{\bm{r}_{p_0}} + \bm{r}\cdot \bm{E} + \bm{r}\bm{r}:\bm{G}+\dots,\label{eqn:uexp}
	\end{align}
	where $\bm{E}=(a_p/L_\Gamma)\nabla \bm{\bar{u}}|_{\bm{r}_{p_0}}$ and $\bm{G}=\frac{1}{2}(a_p^2/L_\kappa^2)\nabla\nabla \bm{\bar{u}}|_{\bm{r}_{p_0}}$ with gradient $L_\Gamma$ and curvature $L_\kappa$ length scales. 
	As a consequence of \eqref{eqn:uexp}, the boundary condition \eqref{eqn:w10bc} is also expanded around  $\bm{r}_{p_0}$, so that in the particle fixed coordinate system
	\begin{align}
		\boldsymbol{w}^{(1)}_0 = \bm{u}_{p_0}-\bm{\bar{u}}  = \bm{u}_{p_0} -\bm{\bar{u}}|_{\bm{r}_{p_0}} - \bm{r}\cdot \bm{E} - \bm{r}\bm{r}:\bm{G}+ \dots \quad \text{on} \quad\bm{r}=1\,, \label{w10genbc}
	\end{align}
	where we have retained the first three terms in the background flow velocity expansion. Owing to the linearity of the leading order unsteady Stokes equation, the solution can generally be expressed as \cite{landau1959course,pozrikidis1992boundary}
	\begin{align}
		\boldsymbol{w}^{(1)}_0  =  \bm{\mathcal{M}}_D \cdot \bm{u}_s  - \bm{\mathcal{M}}_Q \cdot\left(\bm{r}\cdot \bm{E}\right) -\bm{\mathcal{M}}_O \cdot\left(\bm{r}\bm{r}:\bm{G}\right) + \dots, \label{w10gen}
	\end{align}
	where $\bm{\mathcal{M}}_{D,Q,O}(r,\lambda)$ are spatially dependent mobility tensors. For oscillatory flows, they depend on the Stokes number $\lambda$. More explicit forms of these tensors will be given below.
	
	With the leading-order disturbance flow field known, the equations at $\mathcal{O}(\operatorname{Re}_p)$ are as follows,
	\begin{subequations}
		\begin{align}
			\nabla^{2} \boldsymbol{w}^{(1)}_1-\nabla p^{(1)}_1 &=\nabla \cdot \bm{\sigma}^{(1)}_1=3\lambda \frac{\partial \boldsymbol{w}^{(1)}_1}{\partial t}+\bm{f}_0, \\
			\boldsymbol{\nabla} \cdot \boldsymbol{w}^{(1)}_1 &=0, \\
			\boldsymbol{w}^{(1)}_1 &= \bm{u}_{p_1} \quad \text { on } \bm{r}=1, \\
			\boldsymbol{w}^{(1)}_1 &= 0 \quad \text { as } \bm{r} \rightarrow \infty \,,
		\end{align}\label{orep}\nolinebreak
	\end{subequations}
	where $\bm{f}_0=\boldsymbol{w}^{(0)}\cdot \nabla \boldsymbol{w}^{(1)}_0+\boldsymbol{w}^{(1)}_0\cdot \nabla \boldsymbol{w}^{(0)} +\boldsymbol{w}^{(1)}_0\cdot \nabla\boldsymbol{w}^{(1)}_0$ is the (explicitly known) leading-order nonlinear forcing of the disturbance flow. In order to compute the force at this order, we employ a reciprocal relation in the Laplace domain since the problem is time-dependent and, for oscillatory flows, the Laplace transform is explicitly obtained.
	\subsection{Reciprocal theorem and test flow}
	A known test flow (denoted by primed quantities such as $\bm{u}'$) is chosen around an oscillating sphere such that it satisfies the following unsteady Stokes equation:
	\begin{subequations}
		\begin{align}
			\nabla^{2} \bm{u}'-\nabla p' &=\nabla \cdot \bm{\sigma}'=3\lambda\frac{\partial \bm{u}'}{\partial t}, \\
			\boldsymbol{\nabla} \cdot \bm{u}' &=0, \\
			\bm{u}' &= u'(t)\,\bm{e} \quad \text { on } \bm{r}=1, \\
			\bm{u}' &= 0 \quad \text { as } \bm{r} \rightarrow \infty,
		\end{align}\label{test}\nolinebreak
	\end{subequations}
	where the unit vector $\bm{e}$ is chosen to coincide with the direction in which the force on the particle is desired. The solution to this problem is of the same form as \eqref{w10gen}, but with only the first term, i.e.,
	\begin{align}
		\bm{u}' = u'(t) \bm{\mathcal{M}}_D \cdot \bm{e}\,.
	\end{align}
	Denoting Laplace transformed quantities by hats (e.g., $\hat{\bm{u}}$), one can write down the following symmetry relation using the divergence theorem (cf. \cite{lovalenti1993hydrodynamic,maxey1983equation,hood2015inertial}):
	\begin{align}
		\oint_S  (\hat{\boldsymbol{w}}^{(1)}_1\cdot \hat{\bm{\sigma}}' - \hat{\bm{u}}'\cdot \hat{\bm{\sigma}}^{(1)}_1)\cdot \bm{m} \, dS= \int_V \left[\nabla \cdot (\hat{\boldsymbol{w}}^{(1)}_1\cdot \hat{\bm{\sigma}}') - \nabla \cdot (\hat{\bm{u}}'\cdot \hat{\bm{\sigma}}^{(1)}_1)\right]dV,
	\end{align}
	where $\bm{m}$ is the outward unit normal vector to the surface (pointing inward over the sphere surface), and $\hat{\bm{\sigma}} = \nabla \hat{\bm{u}} +(\nabla \hat{\bm{u}})^T - \hat{p}\bm{I}$. Substituting boundary conditions from \eqref{orep} and \eqref{test}, and setting the volume equal to the fluid-filled domain, we obtain
	\begin{align}
		&\hat{\bm{u}}_{p_1}^{(1)} \cdot \int_{S_p} ( \hat{\bm{\sigma}}'\cdot \bm{m})dS - \hat{u}' \bm{e} \cdot \int_{S_p} ( \hat{\bm{\sigma}}^{(1)}_1\cdot \bm{m})dS +  \int_{S_\infty} ( \hat{\boldsymbol{w}}^{(1)}_1 \cdot\hat{\bm{\sigma}}')\cdot \bm{m}dS - \int_{S_\infty} ( \hat{\bm{u}}' \cdot\hat{\bm{\sigma}}^{(1)}_1)\cdot \bm{m}dS \nonumber\\
		=& \int_V \left[\hat{\boldsymbol{w}}^{(1)}_1 \cdot (\nabla\cdot \hat{\bm{\sigma}}') - \hat{\bm{u}}' \cdot (\nabla\cdot \hat{\bm{\sigma}}^{(1)}_1) + \nabla\hat{\boldsymbol{w}}^{(1)}_1 :\hat{\bm{\sigma}}' - \nabla\hat{\bm{u}}' :\hat{\bm{\sigma}}^{(1)}_1 \right]dV\,.
	\end{align}
	The third term on the LHS is $0$ since the viscous test flow stress tensor decays to zero at infinity. Similarly, the integral in the fourth term vanishes in the far field if viscous stresses dominate inertial terms, and also in the case of inviscid irrotational flows (see \cite{lovalenti1993hydrodynamic,stone2001inertial}). The third and fourth terms on the RHS also go to zero, owing to incompressibilty and symmetry of the stress tensor:
	\begin{align}
		&\nabla\hat{\boldsymbol{w}}^{(1)}_1 :\hat{\bm{\sigma}}' - \nabla\hat{\bm{u}}' :\hat{\bm{\sigma}}^{(1)}_1\nonumber\\
		&= \nabla\hat{\boldsymbol{w}}^{(1)}_1 :(\nabla\hat{\bm{u}}'+(\nabla\hat{\bm{u}}')^T)-\hat{p}' \nabla \cdot \hat{\boldsymbol{w}}^{(1)}_1- \nabla\hat{\bm{u}}' : (\nabla\hat{\boldsymbol{w}}^{(1)}_1+(\nabla\hat{\boldsymbol{w}}^{(1)}_1)^T)-\hat{p}^{(1)} \nabla \cdot \hat{\bm{u}}'=0\,.
	\end{align}
	The divergence of the hatted stress tensors in the remaining two terms of the RHS can be obtained by taking the Laplace transforms of \eqref{orep} and \eqref{test} and using the property $\widehat{f'(t)} = s\widehat{f(t)}-f(0)$, so that
	\begin{subequations}
		\begin{align}
			\nabla\cdot \hat{\bm{\sigma}}' &= \bar{\lambda} s \hat{\bm{u}}' - \bm{u}'(0),\\
			\nabla\cdot \hat{\bm{\sigma}}^{(1)}_1 &= \bar{\lambda} s \hat{\boldsymbol{w}}^{(1)}_1 - \boldsymbol{w}^{(1)}_1(0) + \hat{\bm{f}}_0\,.
		\end{align}
	\end{subequations}
	Now, the force on the sphere at this order is given by $\bm{F}^{(1)}_1 =\int_{S_p} ( \bm{\sigma}^{(1)}_1\cdot \bm{n})dS=-\int_{S_p} ( \bm{\sigma}^{(1)}_1\cdot \bm{m})dS$, since $\bm{m}$ points inwards while $\bm{n}$ points outwards on the surface of the sphere. Assuming both flows start from rest, we have (cf. \cite{lovalenti1993hydrodynamic})
	\begin{align}
		\hat{u}' \bm{e}\cdot \frac{\hat{\bm{F}}^{(1)}_1}{F_S/(6\pi)} &= \hat{\bm{u}}_{p_1} \cdot \int_{S_p} ( \hat{\bm{\sigma}}'\cdot \bm{n})dS - \int_V \hat{\bm{u}}' \cdot \hat{\bm{f}}_0dV +\mathcal{O}(\operatorname{Re}_p^2)\,.
	\end{align}
	Adding the force contribution from the previous order, the net force on the particle due to its disturbance flow  reads
	\begin{subequations}
		\begin{align}
			\hat{u}' \bm{e}\cdot \frac{\hat{\bm{F}}^{(1)}}{F_S/(6\pi)} &=\hat{u}' \bm{e}\cdot \left(\hat{\bm{F}}^{(1)}_0+\operatorname{Re}_p\hat{\bm{F}}^{(1)}_1\right) +\mathcal{O}(\operatorname{Re}_p^2)\nonumber\\
			&=\int_{S_p}\left(\hat{\bm{u}}_{p_0}-\hat{\bar{\bm{u}}}+\operatorname{Re}_p\hat{\bm{u}}_{p_1}\right) \cdot  ( \hat{\bm{\sigma}}'\cdot \bm{n})dS - \operatorname{Re}_p\int_V \hat{\bm{u}}' \cdot \hat{\bm{f}}_0dV +\mathcal{O}(\operatorname{Re}_p^2)\\
			\implies \bm{e}\cdot \bm{F}^{(1)} &= \frac{F_S}{6\pi}\mathcal{L}^{-1}\left\{\int_{S_p}\frac{\left(\hat{\bm{u}}_{p}-\hat{\bar{\bm{u}}}\right)}{\hat{u}'}\cdot ( \hat{\bm{\sigma}}'\cdot \bm{n})dS -  \frac{1}{\hat{u}'}\operatorname{Re}_p\int_V \hat{\bm{u}}' \cdot \hat{\bm{f}}_0dV\right\} +\mathcal{O}(\operatorname{Re}_p^2),\label{eqn:F11laplace}
		\end{align}
	\end{subequations}
	where we have used $\bm{u}_p=\bm{u}_{p_0}+\operatorname{Re}_p \bm{u}_{p_1}+\mathcal{O}(\operatorname{Re}_p^2)$, and  $\mathcal{L}^{-1}$ denotes the inverse Laplace transform. The first term on the RHS of \eqref{eqn:F11laplace} is denoted as $F^{(1)}_0$ in the main text (and is the same as that obtained by MR), while the second term represents the $\mathcal{O}(\operatorname{Re}_p)$ inertial force and is denoted as $F^{(1)}_1$ in the main text.
	
	\section{Evaluation of the $\mathcal{O}(\operatorname{Re}_p)$ inertial force}
	In this section, we will explicitly evaluate the volume integral  in \eqref{eqn:F11laplace} representing the $\mathcal{O}(\operatorname{Re}_p)$ inertial force. This requires obtaining $\bm{f}_0$ from the leading-order oscillatory disturbance flow field $\boldsymbol{w}^{(1)}_0$.
	
	\subsection{General solution to equation \eqref{eqn:w01}}
	
	We already remarked that, given the background flow field expansion in uniform, linear, and quadratic parts around the particle,  $\boldsymbol{w}^{(1)}_0$ is formally obtained as the linear combination \eqref{w10gen}. For harmonically oscillating, axisymmetric background flows (i.e., $\bm{\bar{u}}(\bm{r})=\{\bar{u}_r,\bar{u}_\theta,0\}$ in the spherical particle coordinate system, with all components $\propto e^{i t}$), general explicit expressions can be derived for the mobility tensors $\bm{\mathcal{M}}_{D,Q,O}$, ensuring no-slip boundary conditions on the sphere order-by-order. A procedure obtaining $\bm{\mathcal{M}}_D$ is described in Landau--Lifshitz \cite{landau1959course}; the other tensors are determined analogously. Using components in spherical coordinates, they read
	\begin{align}
		\bm{\mathcal{M}}_D = \begin{bmatrix}
			\frac{2a(r)}{r^2}  & 0 & 0 \\
			0 & \frac{a'(r)}{r} & 0 \\
			0 & 0 & 0
		\end{bmatrix}, \quad
		\bm{\mathcal{M}}_Q = \begin{bmatrix}
			\frac{b(r)}{r^3}  & 0 & 0 \\
			0 & \frac{b'(r)}{3r^2} & 0 \\
			0 & 0 & 0
		\end{bmatrix}, \quad
		\bm{\mathcal{M}}_O = \begin{bmatrix}
			\frac{-32c(r)}{3r^4}  & 0 & 0 \\
			0 & \frac{8c'(r)}{3r^3} & 0 \\
			0 & 0 & 0
		\end{bmatrix},
	\end{align}
	where
	\begin{subequations}
		\begin{align}
			a(r)&=\frac{1}{2\beta^2 r} \left[\beta^2 - 3i\beta+3 - 3 e^{-i\beta(r-1)}\left(1+i\beta r\right) \right],\\
			b(r)& = \frac{1}{\beta^2 (\beta-i) r^2}\left[\beta  (-15+\beta  (\beta -6 i))+15 i +5 e^{-i \beta  (r-1)} (\beta  r (3+i \beta  r)-3 i)\right],\\
			c(r)&=\frac{-3 (105+\beta  (\beta  (-45+\beta  (\beta -10 i))+105 i))+ 21 e^{-i \beta  (r-1)} (15+\beta  r (-\beta  r (6+i \beta  r)+15 i))}{32 \beta ^2 (-3+\beta  (\beta -3 i)) r^3},
		\end{align}
	\end{subequations}
	and $\beta = \sqrt{-ia_p^2/(\nu/\omega)}=\sqrt{-3i\lambda}$ is the complex oscillatory boundary layer thickness. We emphasize that these expressions are the same for arbitrary axisymmetric oscillatory $\bm{\bar{u}}$. Accordingly, only the expansion coefficients $\bm{u}_s $, $\bm{E}$, and $\bm{G}$ contain information about the particular flow.
	
	Similarly, the solution to the unsteady test flow is obtained directly as
	\begin{align}
		\bm{u}'=\bm{\mathcal{M}}_D \cdot \begin{bmatrix}
			\cos \theta \\
			-\sin \theta \\
			0
		\end{bmatrix}e^{it}\,. \label{utest}
	\end{align}
	It is understood everywhere that physical quantities are obtained by taking real parts of these complex functions.

	\subsection{Evaluation of $F^{(1)}_1$}
	In order to compute the volume integral in \eqref{eqn:F11laplace}, we first note that only certain products in $\bm{f}_0$ are non-vanishing when the angular integration over $\theta$ is performed. In particular, due to alternating symmetry of terms in the background flow field expansion \eqref{eqn:uexp}, and consequently in the leading order disturbance flow \eqref{w10gen}, only products of adjacent terms survive. This is because, in the Taylor expansion of the background flow field, the first and third terms are symmetric ($\bm{u}(-\bm{r})=\bm{u}(\bm{r})$) while the second one is anti-symmetric ($\bm{u}(-\bm{r})=-\bm{u}(\bm{r})$). For example, the first term in $\bm{f}_0$ reads
	\begin{align}
		\boldsymbol{w}^{(0)} \cdot \nabla \boldsymbol{w}^{(1)}_0
		= \left(-\bm{u}_s  + \bm{r}\cdot \bm{E}+ \bm{r}\bm{r}:\bm{G} \right)\cdot \nabla\left(\bm{\mathcal{M}}_D \cdot \bm{u}_s  - \bm{\mathcal{M}}_Q \cdot\left(\bm{r}\cdot \bm{E}\right) -\bm{\mathcal{M}}_O \cdot\left(\bm{r}\bm{r}:\bm{G}\right) \right),
	\end{align}
	and the only terms that survive the angular integration are the symmetric ones (after a contraction with the symmetric test flow $\bm{u}'$), i.e.,
	\begin{align}
		\left(-\bm{u}_s + \bm{r}\bm{r}:\bm{G} \right)\cdot \nabla\left(-\bm{\mathcal{M}}_Q\cdot \left(\bm{r}\cdot\bm{E}\right) \right) +\left( \bm{r}\cdot \bm{E} \right)\cdot \nabla\left(\bm{\mathcal{M}}_D \cdot \bm{u}_s  -\bm{\mathcal{M}}_O \cdot\left(\bm{r}\bm{r}:\bm{G}\right) \right)\,.
	\end{align}
	Furthermore, in this paper we restrict ourselves to the case of neutrally buoyant particles and consequently the slip velocity is $\bm{u}_s=0$. In summary, only the following terms in $\bm{f}_0$ have non-trivial contributions to the volume integral:
	\begin{align}
		\bm{f}_0 =& -(\bm{r}\bm{r}:\bm{G})\cdot \nabla\left(\bm{\mathcal{M}}_Q\cdot (\bm{r}\cdot\bm{E})\right) - \left( \bm{r}\cdot \bm{E} \right)\cdot \nabla\left(\bm{\mathcal{M}}_O \cdot\left(\bm{r}\bm{r}:\bm{G}\right) \right)\nonumber\\
		&-\left(\bm{\mathcal{M}}_Q\cdot (\bm{r}\cdot\bm{E})\right)\cdot\nabla(\bm{r}\bm{r}:\bm{G}) - \left(\bm{\mathcal{M}}_O \cdot\left(\bm{r}\bm{r}:\bm{G}\right) \right)\cdot \nabla\left( \bm{r}\cdot \bm{E} \right)\nonumber\\
		&+ \left(\bm{\mathcal{M}}_Q\cdot (\bm{r}\cdot\bm{E})\right)\cdot\nabla\left(\bm{\mathcal{M}}_O \cdot\left(\bm{r}\bm{r}:\bm{G}\right) \right)+\left(\bm{\mathcal{M}}_O \cdot\left(\bm{r}\bm{r}:\bm{G}\right) \right)\cdot \nabla\left(\bm{\mathcal{M}}_Q\cdot (\bm{r}\cdot\bm{E})\right). \label{f0contribs}
	\end{align}
	All information about the background flow field is contained in the constant tensors $\bm{E}$ and $\bm{G}$, which are evaluated at the particle position. If the particle is farther away from the surface of the oscillating object exciting the flow than the Stokes layer thickness $\delta_S$, it is exposed to a pure potential flow; this will be the case in the overwhelming majority of realistic scenarios. For potential flows it can be shown that all non-zero terms of \eqref{f0contribs} are proportional to $\bm{E}:\bm{G}$. Choosing a test flow in direction $\bm{e}$, one obtains a surprisingly compact result for the $\bm{e}$-component of the $\mathcal{O}(\operatorname{Re}_p)$ inertial force:
	\begin{align}
		\left\langle\frac{F^{(1)}_1}{F_S}\right\rangle=-\frac{1}{6\pi}\bigg\langle\mathcal{L}^{-1}\left\{\frac{1}{\hat{u}' }\int_V \hat{\bm{u}}'\cdot \hat{\bm{f}}_0dV\right\}\bigg\rangle=\frac{4}{9}\langle \bm{E}:\bm{G} \rangle \cdot \bm{e}\, \mathcal{F}_1^{(1)}(\lambda)\,. \label{f11result}
	\end{align}
	We have here applied the required Laplace transforms as well as a time average to extract the steady part of the force. Performing the volume integral leaves a universal dimensionless function $\mathcal{F}(\lambda)$, whose contributions stem from $\bm{\mathcal{M}}_{D,Q,O}$. Explicitly, this function reads
	\begin{align}
		& \mathcal{F}_1^{(1)}(\lambda)=\bigg[2\pi \bigg(-796500 \bar{\lambda }^{3/2}-336636 \bar{\lambda }^{5/2}+34005 \bar{\lambda }^{7/2}+59790 \bar{\lambda }^{9/2}+3312 \bar{\lambda }^{11/2}+568 \bar{\lambda }^6\nonumber\\
		&+14078 \bar{\lambda }^5+97470 \bar{\lambda }^4-109920 \bar{\lambda }^3-646137 \bar{\lambda }^2-648594 \bar{\lambda }-322056 \sqrt{\bar{\lambda }}-76545\bigg)\nonumber\\
		&+ e^{(1-i) \sqrt{\bar{\lambda }}} \pi  \bar{\lambda }^{5/2} \bigg(9 \bigg(\pi  (4410+2033 i)-28176 e^{(1+i) \sqrt{\bar{\lambda }}} \text{Ei}\left(-2 \sqrt{\bar{\lambda }}\right)+e^{(2+2 i) \sqrt{\bar{\lambda }}} (5600-12600 i) \text{Ei}\left((-3-i) \sqrt{\bar{\lambda }}\right)\nonumber\\
		&-(2033+4410 i) e^{2 i \sqrt{\bar{\lambda }}} \text{Ei}\left((-1-i) \sqrt{\bar{\lambda }}\right)+e^{2 \sqrt{\bar{\lambda }}} (5600+12600 i) \text{Ei}\left((-3+i) \sqrt{\bar{\lambda }}\right)-(2033-4410 i) \text{Ei}\left((-1+i) \sqrt{\bar{\lambda }}\right)\nonumber\\
		&+e^{(2+2 i) \sqrt{\bar{\lambda }}} (12600+5600 i) \pi +e^{2 i \sqrt{\bar{\lambda }}} (4410-2033 i) \pi +e^{2 \sqrt{\bar{\lambda }}} (12600-5600 i) \pi \bigg) \bar{\lambda }^{3/2}+6 \bigg(\pi  (4195+3982 i)\nonumber\\
		&-28080 e^{(1+i) \sqrt{\bar{\lambda }}} \text{Ei}\left(-2 \sqrt{\bar{\lambda }}\right)-(3982+4195 i) e^{2 i \sqrt{\bar{\lambda }}} \text{Ei}\left((-1-i) \sqrt{\bar{\lambda }}\right)-(3982-4195 i) \text{Ei}\left((-1+i) \sqrt{\bar{\lambda }}\right)\nonumber\\
		&+e^{2 i \sqrt{\bar{\lambda }}} (4195-3982 i) \pi \bigg) \bar{\lambda }^{5/2}+4 \bigg(\pi  (241+1714 i)+720 e^{(1+i) \sqrt{\bar{\lambda }}} \text{Ei}\left(-2 \sqrt{\bar{\lambda }}\right)-(1714+241 i) e^{2 i \sqrt{\bar{\lambda }}} \text{Ei}\left((-1-i) \sqrt{\bar{\lambda }}\right)\nonumber\\
		&-(1714-241 i) \text{Ei}\left((-1+i) \sqrt{\bar{\lambda }}\right)+e^{2 i \sqrt{\bar{\lambda }}} (241-1714 i) \pi \bigg) \bar{\lambda }^{7/2}-(120+120 i) \bigg(\pi  \left(-i+e^{2 i \sqrt{\bar{\lambda }}}\right)-i e^{2 i \sqrt{\bar{\lambda }}} \text{Ei}\left((-1-i) \sqrt{\bar{\lambda }}\right)\nonumber\\
		&+\text{Ei}\left((-1+i) \sqrt{\bar{\lambda }}\right)\bigg) \bar{\lambda }^{9/2}-(4+4 i) \bigg(\pi  \left(e^{2 i \sqrt{\bar{\lambda }}} (248+127 i)+(-127-248 i)\right)+e^{2 i \sqrt{\bar{\lambda }}} (127-248 i) \text{Ei}\left((-1-i) \sqrt{\bar{\lambda }}\right)\nonumber\\
		&+(248-127 i) \text{Ei}\left((-1+i) \sqrt{\bar{\lambda }}\right)\bigg) \bar{\lambda }^4-(6+6 i) \bigg(e^{2 i \sqrt{\bar{\lambda }}} \pi  (567+2134 i)+e^{(1+i) \sqrt{\bar{\lambda }}} (736-736 i) \text{Ei}\left(-2 \sqrt{\bar{\lambda }}\right)\nonumber\\
		&+e^{2 i \sqrt{\bar{\lambda }}} (2134-567 i) \text{Ei}\left((-1-i) \sqrt{\bar{\lambda }}\right)+(567-2134 i) \text{Ei}\left((-1+i) \sqrt{\bar{\lambda }}\right)+(-2134-567 i) \pi \bigg) \bar{\lambda }^3\nonumber\\
		&+\bigg(\pi  (39033+25089 i)-381504 e^{(1+i) \sqrt{\bar{\lambda }}} \text{Ei}\left(-2 \sqrt{\bar{\lambda }}\right)-(25089+39033 i) e^{2 i \sqrt{\bar{\lambda }}} \text{Ei}\left((-1-i) \sqrt{\bar{\lambda }}\right)\nonumber\\
		&-(25089-39033 i) \text{Ei}\left((-1+i) \sqrt{\bar{\lambda }}\right)+e^{2 i \sqrt{\bar{\lambda }}} (39033-25089 i) \pi \bigg) \bar{\lambda }^2+(315+315 i) \bigg(e^{(2+2 i) \sqrt{\bar{\lambda }}} \pi  (420+60 i)\nonumber\\
		&-(96-96 i) e^{(1+i) \sqrt{\bar{\lambda }}} \text{Ei}\left(-2 \sqrt{\bar{\lambda }}\right)+e^{(2+2 i) \sqrt{\bar{\lambda }}} (60-420 i) \text{Ei}\left((-3-i) \sqrt{\bar{\lambda }}\right)-(49+28 i) e^{2 i \sqrt{\bar{\lambda }}} \text{Ei}\left((-1-i) \sqrt{\bar{\lambda }}\right)\nonumber\\
		&+e^{2 \sqrt{\bar{\lambda }}} (420-60 i) \text{Ei}\left((-3+i) \sqrt{\bar{\lambda }}\right)+(28+49 i) \text{Ei}\left((-1+i) \sqrt{\bar{\lambda }}\right)-(60+420 i) e^{2 \sqrt{\bar{\lambda }}} \pi +(49-28 i) \pi \nonumber\\
		&+e^{2 i \sqrt{\bar{\lambda }}} (28-49 i) \pi \bigg) \bar{\lambda }+15120 e^{\sqrt{\bar{\lambda }}} \bigg(-5 i e^{\sqrt{\bar{\lambda }}} \pi +5 i e^{(1+2 i) \sqrt{\bar{\lambda }}} \pi +5 e^{(1+2 i) \sqrt{\bar{\lambda }}} \text{Ei}\left((-3-i) \sqrt{\bar{\lambda }}\right)+5 e^{\sqrt{\bar{\lambda }}} \text{Ei}\left((-3+i) \sqrt{\bar{\lambda }}\right)\nonumber\\
		&-e^{i \sqrt{\bar{\lambda }}} \text{Ei}\left(-2 \sqrt{\bar{\lambda }}\right)\bigg)+945 \bigg(7 \pi +7 e^{2 i \sqrt{\bar{\lambda }}} \pi -160 i e^{2 \sqrt{\bar{\lambda }}} \pi +160 i e^{(2+2 i) \sqrt{\bar{\lambda }}} \pi +160 e^{(2+2 i) \sqrt{\bar{\lambda }}} \text{Ei}\left((-3-i) \sqrt{\bar{\lambda }}\right)\nonumber\\
		&+160 e^{2 \sqrt{\bar{\lambda }}} \text{Ei}\left((-3+i) \sqrt{\bar{\lambda }}\right)-48 e^{(1+i) \sqrt{\bar{\lambda }}} \text{Ei}\left(-2 \sqrt{\bar{\lambda }}\right)-7 i e^{2 i \sqrt{\bar{\lambda }}} \text{Ei}\left((-1-i) \sqrt{\bar{\lambda }}\right)+7 i \text{Ei}\left((-1+i) \sqrt{\bar{\lambda }}\right)\bigg) \sqrt{\bar{\lambda }}\bigg)\bigg]\bigg/ \nonumber\\
		&\bigg[15120 \sqrt{\bar{\lambda }} \left(84 \bar{\lambda }^{3/2}+32 \bar{\lambda }^{5/2}+8 \bar{\lambda }^3+64 \bar{\lambda }^2+72 \bar{\lambda }+36 \sqrt{\bar{\lambda }}+9\right)\bigg].\label{F11ful}
	\end{align}
	Here $\bar{\lambda }=(3/2)\lambda$ and $\text{Ei}$ is the exponential integral function. We show below that this lengthy expression is approximated to great accuracy by two simple terms.
	
	We stress again here that the result is universal for any oscillatory potential flow; for the prototypical case of the volumetrically oscillating bubble,
	$\langle \bm{E}:\bm{G} \rangle\cdot \bm{e}_r = -9/r_{p}^7$, as noted in the Methods section.

	\subsection{Net inertial force}
	The time-averaged force contribution from the background flow at  $\mathcal{O}(\operatorname{Re}_p)$ is of the same form as \eqref{f11result}, except that $\mathcal{F}(\lambda)$ is replaced by the simple constant $\mathcal{F}_1^{(0)}=\frac{1}{5}$ \cite{ral20_MRNote_preprint}. The two contributions $F^{(1)}_1$ and $F^{(0)}_1$ can thus be simply added. Transforming back to dimensional variables, we obtain
	the net time-averaged force on the particle as
	\begin{align} \label{AvgForce}
		\bm{F}_{\Gamma \kappa} = m_f a_p^2 \left< \nabla \bar{\bm{U}} :  \nabla \nabla \bar{\bm{U}}\right> \mathcal{F}(\lambda),
	\end{align}
	where $\mathcal{F} = \mathcal{F}_1^{(1)} + \mathcal{F}_1^{(0)}$ and $m_f = 4 \pi a_p^3/3$, as noted in the main text. This time-averaged inertial force on the particle is derived for a background flow that is symmetric about an axis $\bm{e}$ passing through the center of the particle.
	
	It was remarked above that the simple form of \eqref{AvgForce} is a consequence of the background flow being potential. This can be backed up by symmetry arguments and dimensional analysis for an arbitrary oscillatory background flow that has a harmonic scalar potential, $\bar{\bm{U}} = \nabla \bar{\varphi}$ with $\nabla^2 \bar{\varphi} = 0$. Such a flow is in fact generic since the background flow vorticity decays exponentially outside the Stokes boundary layer of the compact object driving the background flow. We are interested in a time-averaged force on the particle that is (i) quadratic in the oscillation amplitude and (ii) involves contractions of the flow gradient $\nabla \bar{\bm{U}} = \nabla \nabla \bar{\varphi}$ and the flow curvature tensor $\nabla \nabla \bar{\bm{U}} = \nabla \nabla \nabla \bar{\varphi}$. The only dimensionless parameter in the problem not already specified by $\bar{\bm{U}}$ is the Stokes number $\lambda$. Collecting the above statements, the only way to construct the time-averaged force (a vector) from the higher rank tensors  $\nabla \nabla \bar{\varphi}$ and $\nabla \nabla \nabla \bar{\varphi}$ is by their contraction $\nabla \nabla \bar{\varphi} : \nabla \nabla \nabla \bar{\varphi}$. All other combinations are either of insufficient tensor rank or are identically zero (since $\nabla^2 \bar{\varphi} = 0$). See \cite{danilov2000mean} for similar arguments for flows without curvature. Including the correct dimensions, the time averaged force for any oscillatory potential background flow thus has the form
	\begin{align}
		\bm{F}_{\Gamma \kappa} = m_f a_p^2 \left< \nabla \nabla \bar{\varphi} :  \nabla\nabla \nabla \bar{\varphi} \right> \mathcal{F}(\lambda) \,. 
	\end{align}
	Note that although the background flow is irrotational, the disturbance flow has a finite vorticity within the Stokes layer around the particle. Under this general setting there is no requirement of axisymmetry of the background flow, so \eqref{AvgForce} as well as (1) in the main text apply to the generic case of an oscillatory potential flow background, and with the same universal function $\mathcal{F}(\lambda)$.

	\section{Accuracy of the uniformly valid expression for $\mathcal{F}$}
	
\begin{figure}[hbt]
		\centering
		\includegraphics[width=\linewidth]{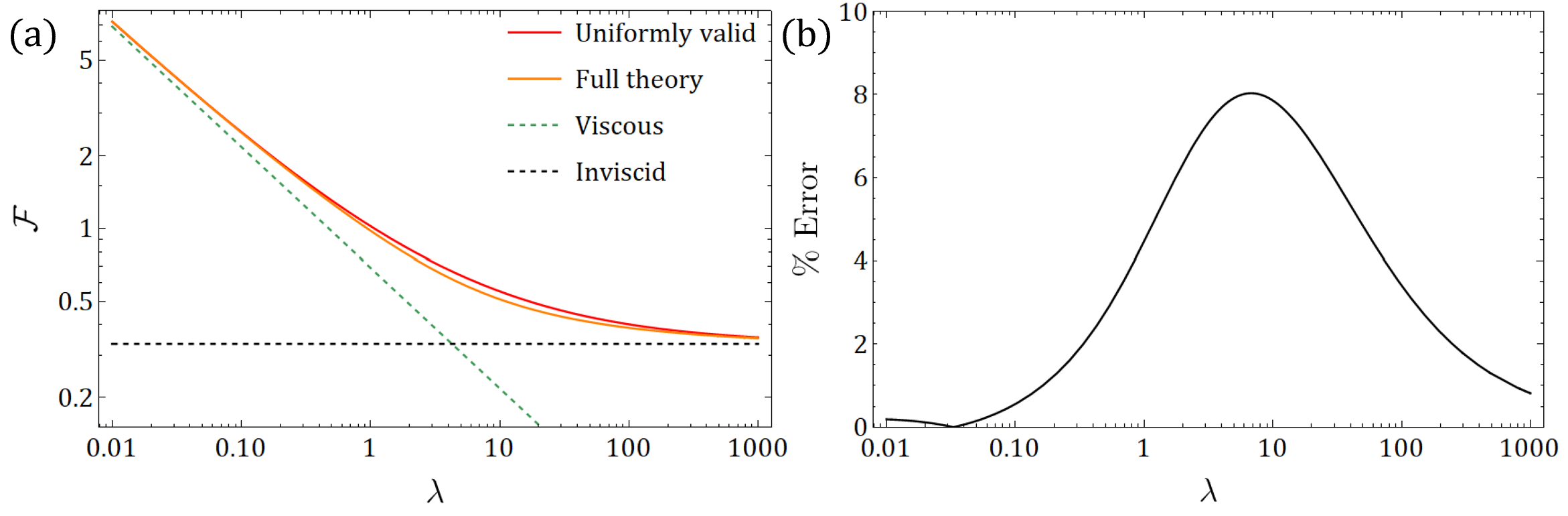}
		\caption{(a) Logarithmic plot of the overall inertial force magnitude ${\mathcal F}(\lambda)$: the uniformly valid expression (red) closely tracks the full solution (orange) while the inviscid theory (gray dashed) severely underestimates the inertial force even for moderately large $\lambda$. (b) The magnitude of the percentage error between the uniformly valid and full solutions is small throughout the entire range of $\lambda$, with a maximum error of $\sim8\%$ where the two limits blend, as one would expect.}
		\label{fig:figuv}
	\end{figure}

	As stated in the main text, while the explicit functional form \eqref{F11ful} of $\mathcal{F}_1^{(1)}(\lambda)$ is rather lengthy, we Taylor expand in both the viscously dominated limit ($\lambda \to 0$) and the inviscid limit ($\lambda \to \infty$) to obtain
	\begin{align}
		\mathcal{F}^{v} = \frac{9}{16}\sqrt{\frac{3}{2\lambda}}+\mathcal{O}(1),\quad \mathcal{F}^{i}=\frac{1}{3} + \mathcal{O}(1/\sqrt{\lambda}).
		\label{Flamexp}
	\end{align}
	We construct a uniformly valid solution by simply adding the two leading order results, yielding $\mathcal{F}(\lambda) = \frac{1}{3}+\frac{9}{16}\sqrt{\frac{3}{2\lambda}}$. In Fig.~\ref{fig:figuv}(a), we plot the uniformly valid $\mathcal{F}$ (red curve) and the full theory represented by Eq.~\eqref{F11ful} (orange), along with the viscous and inviscid limits denoted by dashed lines. Figure~\ref{fig:figuv}(b) shows that the relative error between the red and orange curves is small ($\lesssim 8\%$) for all $\lambda$, even those far smaller or larger than practically relevant values.
	
	\section{Fitting procedure to obtain $\mathcal{F}$ from DNS}
	The DNS outputs (unsteady) particle trajectories as a function of time, with an initial transient period when the particle starts from rest before periodic motion is fully established. As shown in Fig.~3(a) of the main text, these oscillatory trajectories were time-averaged to obtain the steady particle dynamics $r_{p}(T)$, which is a function of the \textit{slow} time $T=\epsilon^2 t$. We fit these trajectories to Eq.~(9) in the main text with $\cal{F}$ as the fitting parameter in order to obtain the simulation points of Fig.~4 of the main text. This was done in two ways: i) Taking a derivative with respect to time of the time-averaged trajectories from simulations, one obtains $\cal{F}$ directly from Eq.~(9) of the main text. ii) The explicit analytical solution to Eq.~(9) of the main text, namely,  $r_{p}(T)=\left(r_{p}(0)-48\alpha^2 T \cal{F} \right)^{1/8}$,  was fitted to the time-averaged trajectories from DNS using the method of least squares over a time interval that excludes the period of initial transient behavior in simulations. We found that both these strategies yielded virtually identical values for $\cal{F}$, which are displayed in Fig.4 of the main text.
	
	\section{Simulation methods and details}
	In order to perform three dimensional flow--structure interaction simulations with deforming geometries, we use the remeshed Vortex Method (rVM) described in \cite{gazzola2011simulations}. Here, we list our simulation methodology and parameters for completeness and reproducibility, as well as convergence tests used to assess simulations accuracy.
	
	\subsection{Fluid--structure interaction}
	We briefly recap the governing equations and numerical method used for our simulation. We consider incompressible viscous flows in an unbounded domain in which two density-matched spherical bodies (i.e. bubble and particle) are immersed.
	We denote the computational domain as $\Omega = \Omega_\text{f} \cup \Omega_\text{B}$, where $\Omega_\text{f}$ is the fluid domain and $\Omega_\text{B} = \Omega_\text{b} \cup \Omega_\text{p}$ is the domain in which the bubble ($\Omega_\text{b}$) and particle ($\Omega_\text{p}$) reside, and denote the interface between the fluid and the bodies as $\partial \Omega_\text{B}$.
	Both the bubble and the particle are then represented by mollified characteristic functions $\chi_\text{b}(\bm{x})$ and $\chi_\text{p}(\bm{x})$, respectively, on a regular Cartesian grid mesh such that $\chi_\text{b}(\bm{x}) = 1$ for $\bm{x} \in \Omega_\text{b}$, $\chi_\text{p}(\bm{x}) = 1$ for $\bm{x} \in \Omega_\text{p}$, and $\chi_\text{b}(\bm{x}) = \chi_\text{p}(\bm{x}) = 0$ for $\bm{x} \in \Omega_\text{f}$.
	In order to avoid discontinuities,
	for each of the bodies, we smoothly blend the $\chi$ values using the mollification function
	\begin{equation} \label{eq:characteristicFunction}
		\chi(d) =   \left\{
		\begin{array}{ll}
			0 & d < -\epsilon_\text{m}, \\
			\frac{1}{2}(1+ \frac{d}{\epsilon_\text{m}} + \frac{1}{\pi}\sin(\pi \frac{d}{\epsilon_\text{m}})) & |d| \leq \epsilon_\text{m} ,\\
			1 & d > \epsilon_\text{m}, \\
		\end{array}
		\right.
	\end{equation}
	where $d$ is the signed-distance to the body--fluid interface, and $\epsilon_\text{m}$ is a user-defined smoothing parameter.
	
	We then solve the incompressible Navier--Stokes equation \cref{eq:ns-vortvel} in its velocity--vorticity form
	\begin{equation}\label{eq:ns-vortvel}
		\nabla \cdot \bm{u} = 0, \quad \frac{D\boldsymbol{\omega}}{Dt} = (\boldsymbol{\omega} \cdot \nabla)\bm{u} + \nu \nabla^2 \boldsymbol{\omega} + \lambda_\text{penal} \nabla \times (\chi(\bm{u}_\text{B}-\bm{u})) ~~~\bm{x}\in\Omega
	\end{equation}
	where $\boldsymbol{\omega}$ is the vorticity field, $\bm{u}$ is the fluid velocity field, $\bm{u}_\text{B}$ is the body velocity and $\nu$ is the kinematic viscosity.
	Here $\lambda_\text{penal} \gg 1$ is the penalization parameter and $\lambda_\text{penal} \nabla \times (\chi(\bm{u}_\text{B}-\bm{u}))$ is the Brinkmann penalization term used to approximate the no-slip boundary condition \cite{gazzola2011simulations}.
	We note that while a bubble has a no-stress boundary condition at the interface, which in general is different from no-slip, in the case of a bubble oscillating in pure breathing mode, where tangential boundary conditions have no effect, both would result in the same flow response.
	Indeed, using this method, fluid velocity within a body is forced to approach the body velocity (i.e. $\bm{u}(\bm{x}) = \bm{u}_\text{B}(\bm{x})$ for $\bm{x} \in \Omega_\text{B}$). The body velocity $\bm{u}_\text{B}$ can be decomposed into its rigid components of motion and the body deformation velocity field as
	$\bm{u}_\text{B}(\bm{x}, t) = \bm{u}_\text{T}(t) + \bm{u}_\text{R}(\bm{x},t) + \bm{u}_\text{def}(\bm{x},t)$,
	where $\bf{u}_\text{T}$ and $\bf{u}_\text{R}$
	are rigid translational and rotational velocities,
	and $\bf{u}_\text{def}$ is the (imposed) deformation velocity field. The body rigid velocity (as a result of action from the fluid) is obtained via a projection approach \cite{gazzola2011simulations} where $\bf{u}_\text{T}$ and $\bf{u}_\text{R}$ are computed through the conservation of momentum in the system.
	The imposed deformation velocity field used to prescribe the bubble's breathing mode is $\bm{u}_\text{def}(\bm{x},t) = \frac{\bm{x}}{a_b} \epsilon a_b \omega \sin(\omega t)$ for $\bm{x} \in \Omega_\text{B}$.
	This methodology based on remeshed vortex methods, penalization and projection has been validated across a range of fluid--structure interaction problems involving both rigid and deformable bodies, from bluff body flows to biological swimming \cite{gazzola2011simulations,gazzola2012flow,gazzola2012c,gazzola2014reinforcement,gazzola2016learning}.
	Recently, it has also been demonstrated in resolving the spatio-temporal scales related to oscillatory flow problems, particularly in viscous streaming settings involving individual and multiple arbitrary-shaped bodies, both in two and three dimensions \cite{parthasarathy2019streaming,bhosale_parthasarathy_gazzola_2020}.
	For a more detailed description of the numerical method, we refer to \cite{gazzola2011simulations}.

	\subsection{Simulation details}
	We simulate both the bubble and the particle as spheres of radii $a_b = 0.01$ and $a_p = 0.002$, respectively (so that $a_p / a_b = 0.2$), and set the mollification smoothing parameter $\epsilon_\text{m} = \sqrt{2} \Delta x$ used in the characteristic function \cref{eq:characteristicFunction}, where $\Delta x$ is the simulation grid size.
	The computational domain is initialized with a physical size of $[-2, 5.875]~a_b \times [-1.9125, 1.9125]~a_b \times [-1.9125, 1.9125]~a_b$. We then discretized the domain with a mesh of N = $560 \times 272 \times 272$ nodes, resulting in a uniform grid size of $\Delta x = 1.40625 \times 10^{-4} = 1.40625 \times 10^{-2}a_b$ in each direction.
	The domain boundary conditions are set to free-space (unbounded) boundary conditions so that $\bm{u} \to 0$ as $\bm{x} \to \infty$. We initialize the particle at $[r_{p}(0), 0, 0]$ and the bubble at $[0, 0, 0]$, both being at rest. While the particle is free to move as a result of its interaction with the fluid, we pin the position of the bubble in place and set the bubble to oscillate in a pure volume (breathing) mode. This is achieved by prescribing $\bm{u}_\text{def}(\bm{x},t) = \frac{\bm{x}}{a_b} \epsilon a_b \omega \sin(\omega t)$ for $\bm{x} \in \Omega_\text{B}$, so that the interface of the bubble moves with radial velocity $\epsilon a_b \omega \sin(\omega t)$. Throughout this paper, we set $\omega = 16\pi$. 
	The viscosity $\nu$ is set based on $\lambda = a_p^2 \omega / (3\nu)$ and the simulation is allowed to run until the particle's steady velocity is achieved (typically 40--200 oscillation cycles, depending on $\lambda$ where larger $\lambda$ require longer time for transient effects to vanish). Finally, we note that the bubble, particle and fluid are density-matched and density is set to unity.

	\subsection{Implementation and resources}
	The simulation algorithm is implemented in Fortan90 and relies on MPI for distributed memory parallelism. The software relies on Parallel Particle Mesh library (PPM) \cite{sbalzarini2006ppm} which provides a convenient abstraction layer over MPI particle--mesh operations, mapping on processors, processor communication and load-balancing. The software also uses FFTW3 library for Poisson solves and HDF library for data output, visualization and post-processing. The simulations performed in this paper typically run for 48--96 hours on 16 nodes, each with 64 threads, on the Stampede2 supercomputing facility.

	\subsection{Resolution convergence test} \label{sec:resolutionTest}
	It is important that we capture the different length scales involved in order to properly resolve the physics at play. We first identify the different physical (bubble oscillations $\epsilon a_b$, particle oscillations, Stokes boundary layer thickness $\delta_S$) and numerical (mollification length $\epsilon_\text{m} = \sqrt{2}\Delta x$) length scales in this problem. Taking these scales into consideration, we then need to ensure that (i) the oscillations are properly resolved (i.e. $\Delta x < $ oscillation amplitudes) and (ii) $\delta_S$ measured from the bubble interface is not embedded in or under-resolved relative to the mollification region (i.e. $\delta_S > \epsilon_\text{m}$).
	
	\begin{figure*}[h]
		\centering
		\includegraphics[width=\textwidth]{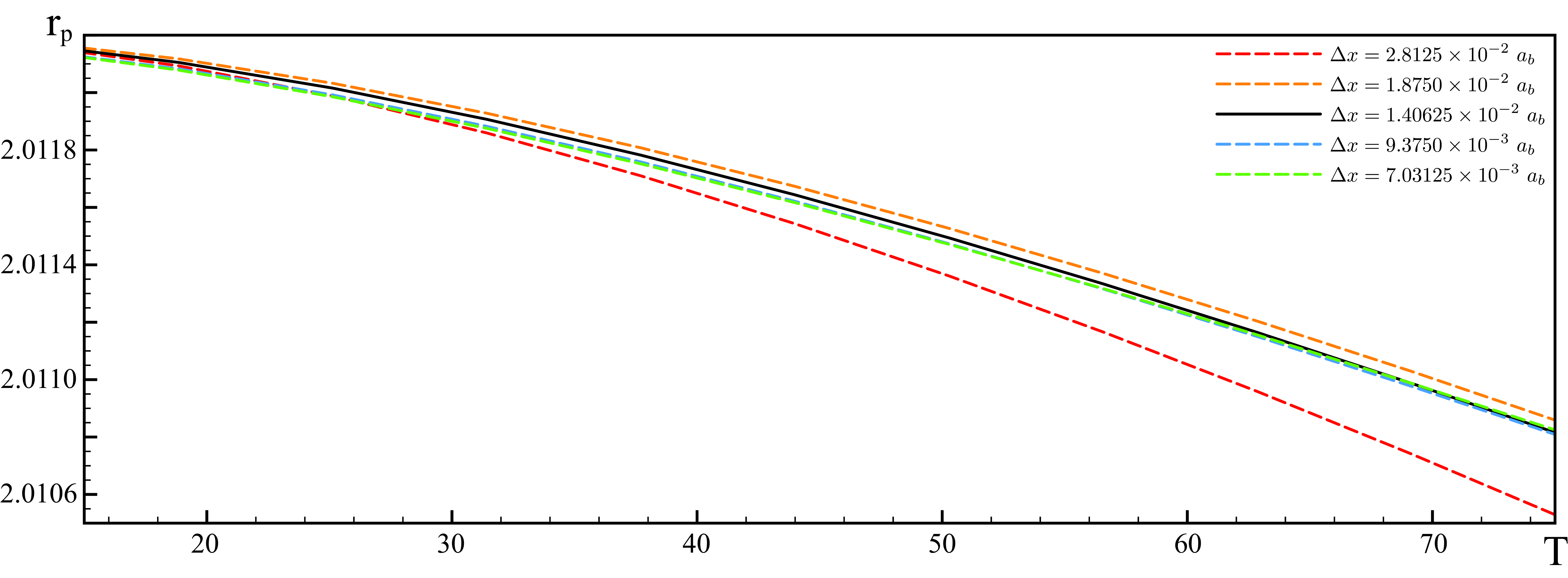}
		\caption{Resolution convergence: Trajectory of particle for simulations with different $\Delta x$ for the case of $r_{p}(0) = 2$ and $\lambda = 20$.}
		\label{fig:resolutionConvergence}
	\end{figure*}
	
	We conduct a resolution convergence test where we run a series of separate simulations with increasing resolution (hence decreasing $\Delta x$).
	We then track the particle's trajectory (via center-to-center distance between bubble and particle) and observe a convergence towards a fixed trajectory, beyond which decreasing the grid size further does not significantly affect the results while requiring considerably larger computational cost.
	We illustrate the convergence behavior in Figure \ref{fig:resolutionConvergence} for the case of $r_{p}(0) = 2$ and $\lambda = 20$, deliberately chosen from the larger $\lambda$ regime in the test cases explored in this paper so that $\delta_S$ is thin (hence requiring finer $\Delta x$ to resolve). Here we note that a grid size of $\Delta x = 1.40625 \times 10^{-2}~a_b$ provides a good compromise between computational cost and accuracy as it resolves the physical and numerical length scales reasonably well, effectively ensuring (i) $\Delta x$ is finer than oscillation amplitudes and (ii) $\delta_S > \epsilon_\text{m}$. Therefore, throughout this paper, we use $\Delta x = 1.40625 \times 10^{-2}~a_b$ for all our simulations.

	\subsection{Domain size convergence test}
	
	\begin{figure*}[h]
		\centering
		\includegraphics[width=\textwidth]{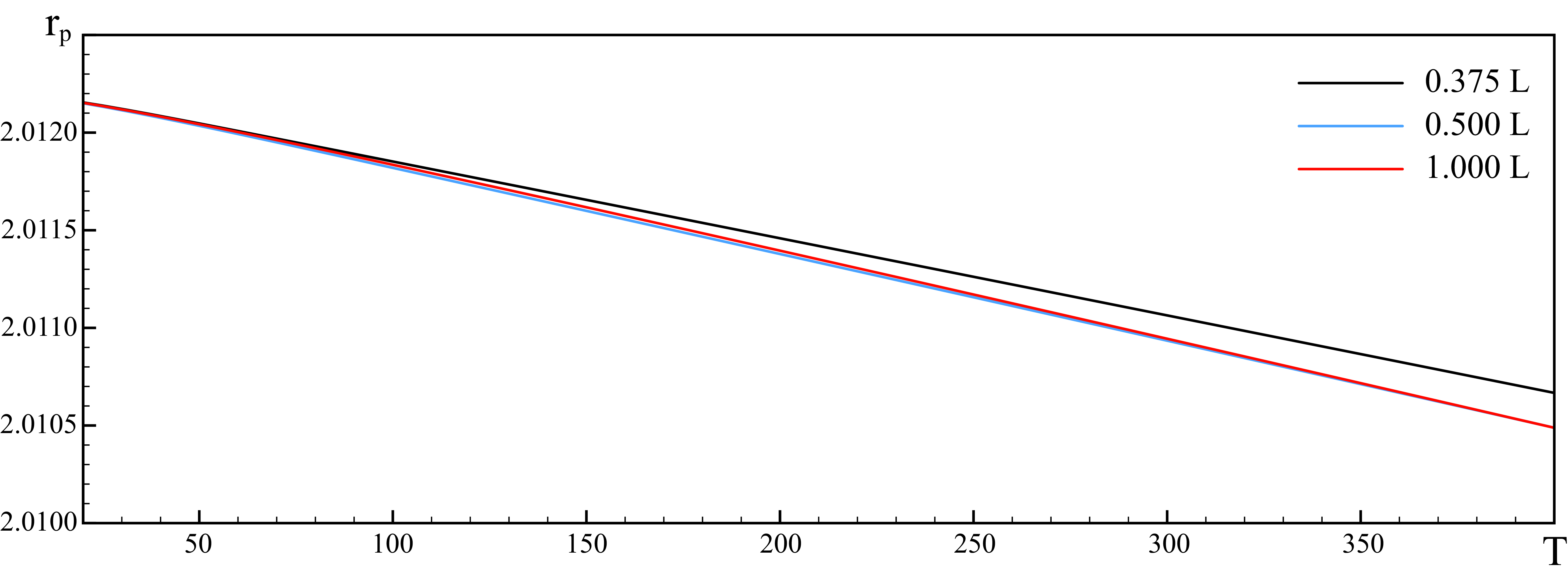}
		\caption{Domain convergence: Trajectory of particle for simulations with different domain size for the case of $r_{p}(0) = 2$ and $\lambda = 1$.}
		\label{fig:domainConvergence}
	\end{figure*}
	
	In order to perform the simulations within feasible computational costs while ensuring that all the length scales involved are properly resolved (see \cref{sec:resolutionTest}), we adjust our simulation domain to a reasonable size such that effects from domain boundaries do not affect the computed results. We perform a simple test by fixing $\Delta x = 2.8125 \times 10^{-2}~a_b$ and explore the boundary effects for different domain sizes. The case of study here is $r_p(0) = 2$ and $\lambda = 1$ (hence a thick $\delta_S$ that might interact with domain boundaries).
	We note that while a lower resolution is used here for exploration (see \cref{sec:resolutionTest}), the $\delta_S$ is still resolved in the simulations (i.e. $\delta_S > \epsilon_\text{m}$) and the test still serves to demonstrate the effects from domain boundaries.
	\Cref{fig:domainConvergence} shows the time-averaged trajectories for different domain sizes 0.375 L, 0.5 L and L, where L = $[-4, 11.75]~a_b \times [-3.825, 3.825]~a_b \times [-3.825, 3.825]~a_b$. We observe that the particle trajectories do not change when doubling the domain size from 0.5 L to L. Therefore, we use the domain size of $[-2, 5.875]~a_b \times [-1.9125, 1.9125]~a_b \times [-1.9125, 1.9125]~a_b$ for simulations conducted throughout this work.

	\bibliography{ms}